\documentclass[10pt]{article}
\pdfoutput=1
\usepackage{amsbsy,amsmath,amsthm,amssymb,graphicx,verbatim,booktabs}

\usepackage{graphicx}

\usepackage{setspace, float, subfig}
\usepackage[longnamesfirst,sort]{natbib}

\usepackage{multirow}
\usepackage{rotating}

\usepackage{geometry}
\theoremstyle{remark}

\newfont{\msbm}{msbm10 at 11pt}
\newcommand {\R} {\mathbb{R}}

\begin{document}
\onehalfspacing

\title{Bayesian inference for a flexible class of bivariate beta distributions}

\author{Roberto C. Crackel \\ Department of Statistics \\ University of California, Riverside \\ {\tt rcrac001@ucr.edu} \and James M. Flegal\footnote{Research supported by the National Science Foundation.}\\ Department of Statistics \\ University of California, Riverside \\ {\tt jflegal@ucr.edu} }  

\date{\today}

\maketitle

\begin{abstract}
Several bivariate beta distributions have been proposed in the literature. In particular, \cite{olkin2003bivariate} proposed a 3 parameter bivariate beta model, which \cite{arnold2011flexible} extend to 5 and 8 parameter models. The 3 parameter model allows for only positive correlation, while the latter models can accommodate both positive and negative correlation. However, these come at the expense of a density that is mathematically intractable. The focus of this research is on Bayesian estimation for the 5 and 8 parameter models. Since the likelihood does not exist in closed form, we apply approximate Bayesian computation, a likelihood free approach. Simulation studies have been carried out for the 5 and 8 parameter cases under various priors and tolerance levels. We apply the 5 parameter model to a real data set by allowing the model to serve as a prior to correlated proportions of a bivariate beta binomial model. Results and comparisons are then discussed.

\end{abstract}

%\begin{keyword}
%% keywords here, in the form: keyword \sep keyword
%Approximate Bayesian computation \sep bivariate beta \sep accept-reject algorithm
%% PACS codes here, in the form: \PACS code \sep code

%% MSC codes here, in the form: \MSC code \sep code
%% or \MSC[2008] code \sep code (2000 is the default)

%\end{keyword}

%\end{frontmatter}

%% \linenumbers

%% main text
\section{Introduction} \label{sec:intro} 
Bivariate beta distributions are becoming increasingly popular across many disciplines.  Furthermore, it is common for Bayesian analysts to use them as prior distributions of correlated binomial random variables. An incomplete list of bivariate beta distributions includes use of the Dirichlet distribution as well as those studied by \cite{arnold2011flexible}, \cite{gupta1985three}, \cite{jones2002multivariate}, \cite{morgenstern1956einfache}, \cite{nadarajah2005some}, \cite{olkin2003bivariate}, \cite{sarmanov1966}, and \cite{ting1996properties}.  \cite{gupta2011non} also consider a non-central bivariate beta model.  An interested reader is directed to \cite{balakrishnan2009continuous} for an extensive list of bivariate beta models along with other bivariate continuous distributions.  

Unfortunately, many bivariate beta models contain parameter and correlation restrictions and hence may not be suitable in applications.  For example, suppose $Z=(Z_1,Z_2)$ defines a bivariate beta random variable.  Then it is well known that if $Z$ follows a Dirichlet distribution the marginals are beta distributed with $z_1+z_2=1$.  Further, the family of bivariate distributions of \cite{morgenstern1956einfache} has a limited correlation range of $(-1/3,1/3)$ as shown by \cite{schucany1978correlation} and those of \cite{olkin2003bivariate} only allow for positive correlation. 

The focus of this paper is on parameter estimation for the flexible 5 and 8 parameter models of \cite{arnold2011flexible}, which extend the 3 parameter specification of \cite{olkin2003bivariate}.  The models of \cite{arnold2011flexible} allow for both positive and negative correlation, that is any correlation in $(-1, 1)$.  The cost of this increased flexibility is a joint density unavailable in closed form, but simulating pseudo-random observations from it is trivial.

The lack of a closed form density eliminates the possibility of maximum likelihood estimation (MLE).  In the 5 parameter model, \cite{arnold2011flexible} propose a clever estimation method, which they refer to as a modified maximum likelihood estimation (MMLE) approach.  This approach uses MLE on the beta distributed marginals to obtain 4 estimating equations.  A final estimating equation is obtained via method of moments using a carefully constructed expectation.  Unfortunately, a simple estimate of this expectation is unstable when a single observation is too close to zero (in one or both dimensions).  Further, there is little hope of finding 4 carefully constructed expectations to extend MMLE to the 8 parameter model.  

This paper proposes a Bayesian approach applicable in the 5 and 8 parameter models, resulting in improved estimation relative to MMLE.  Given that the joint density does not exist in closed form, we cannot compute the posterior distribution with usual techniques.  Instead, we will bypass this difficulty by considering a likelihood free method known as approximate Bayesian computation (ABC). \cite{rubin1984bayesianly} first described elements of the ABC algorithm, however it was \cite{tavare1997inferring} who laid the groundwork for the original ABC algorithms. The basic idea of ABC is to a generate a candidate parameter from the prior distribution and based on this parameter value, an auxiliary data set is generated.  If the auxiliary data is sufficiently ``close'' to the observed data, then the candidate parameter is accepted as a plausible value.  The accepted parameter values via this ABC accept-reject (ABC-AR) algorithm form an i.i.d.\ sample from a distribution that approximates the true posterior, where the approximation is dependent upon the notion of ``close''.

In this paper, we consider an ABC-AR algorithm using various parameter settings, priors, sample sizes, and tolerance levels. We show use of a posterior mean obtained via ABC improves parameter estimation over MMLE in the 5 parameter model. Specifically, our simulations show a significant decrease in mean square error (MSE) even in the presence of bias introduced via a Bayesian approach. The decrease in MSE depends upon the true parameter values and selected prior. Furthermore, we illustrate ABC estimation for the 8 parameter model where we know of no other existing approach. 

Finally, we make application of the 5 parameter bivariate beta model in a bivariate beta binomial context.  Specifically, the bivariate beta binomial distribution is used to model the purchasing habits of bacon and eggs, described previously by \cite{danaher2005bacon}. In this model, we allow the 5 parameter model to serve as a prior distribution to correlated proportions and it serves as a competitor to the \cite{sarmanov1966} bivariate beta model that \cite{danaher2005bacon} proposed. Our analysis considers the ABC-AR algorithm and a Metropolis Hastings based modification proposed by \cite{marjoram2003markov}.

The rest of this paper is organized as follows, Section~\ref{sec:beta} introduces the bivariate beta models of \cite{arnold2011flexible} and MMLE.  Section~\ref{sec:ABC} outlines Bayesian inference for this model along with ABC algorithms for exploring the posterior. Section~\ref{sec:sim} describes the simulation study and discusses findings. In section~\ref{sec:Example}, we describe the application of our 5 parameter model to the bacon and eggs data set. 

\section{Bivariate beta model} \label{sec:beta}

The 8 parameter model of \cite{arnold2011flexible} is defined as follows.  Suppose $U_{i} {\sim} \Gamma(\delta_i,1)$ for $i=1,\dots,8$ and let
\[
V_1=\frac{U_1+U_5+U_7}{U_3+U_6+U_8} \text{ and } V_2=\frac{U_2+U_5+U_8}{U_4+U_6+U_7} \; .
\]
If $Z_1= V_1 / (1+V_1)$ and $Z_2=V_2 / (1+V_2)$, then $Z=(Z_1,Z_2)$ defines a bivariate beta random variable, which we will denote as $\mathcal{B}\mathcal{B}(\delta_1,\dots,\delta_8)$.  It is easy to show the marginal distributions of $Z_1$ and $Z_2$ are beta distributed, i.e.\
\[
Z_1 {\sim} Beta(\delta_{1}+\delta_{5}+\delta_{7},\delta_{3}+\delta_{6}+\delta_{8}) \text{ and } Z_2 {\sim} Beta(\delta_{2}+\delta_{5}+\delta_{8},\delta_{4}+\delta_{6}+\delta_{7}) \; .
\]
The 8 parameter model reduces to the 5 parameter model when ${\delta_3}={\delta_4}={\delta_5}=0$ and by setting ${\alpha_1}={\delta_1}$, ${\alpha_2}={\delta_2}$, ${\alpha_3}={\delta_7}$, ${\alpha_4}={\delta_8}$, and ${\alpha_5}={\delta_6}$. We denote the 5 parameter model as $\mathcal{B}\mathcal{B}(\alpha_1,\dots,\alpha_5)$. Thus, for the 5 parameter model, we have $Z_1 {\sim} Beta(\alpha_{1}+\alpha_{3},\alpha_{4}+\alpha_{5})$ and $Z_2 {\sim} Beta(\alpha_{2}+\alpha_{4},\alpha_{3}+\alpha_{5})$. The 5 parameter model reduces to the 3 parameter model of \cite{olkin2003bivariate} with parameters ${\alpha_1},{\alpha_2}$ and ${\alpha_5}$ by setting ${\alpha_3}={\alpha_4}=0$.

Under a similar construction, one can construct a $k$-variate flexible beta distribution with $2k+1$ parameters \citep{arnold2011flexible}. An interested reader is directed to \cite{olkin2015constructions} for other constructions of $k$-variate beta distributions.

\subsection{Modified maximum likelihood estimation}

Since the joint density does not exist for the 5 and 8 parameter models, standard estimation techniques are unavailable.  In the 5 parameter setting, \cite{arnold2011flexible} propose 3 estimation techniques. The most promising of these is MMLE, which combines MLE for the marginals of $Z_1$ and $Z_2$ with a method of moments estimate.

Suppose we have $n$ observations from the 5 parameter bivariate beta model. Specifically, define $\vec{\boldsymbol{z}}_1 = (z_{11},z_{21},\ldots,z_{n1})^\prime$, $\vec{\boldsymbol{z}}_2 = (z_{12},z_{22},\ldots,z_{n2})^\prime$, and $\tilde{\boldsymbol{z}} = (\vec{\boldsymbol{z}}_1,\vec{\boldsymbol{z}}_2)$. Based on the marginals of $Z_1$ and $Z_2$, the MLE for $a={\alpha_1}+{\alpha_3}$, $b={\alpha_4}+{\alpha_5}$, $c={\alpha_2}+{\alpha_4}$, and $d={\alpha_3}+{\alpha_5}$ can be easily obtained.  We denote the MLE by ${\hat{a}}$, ${\hat{b}}$, ${\hat{c}}$, and ${\hat{d}}$, respectively.  

Further, one can show
\begin{align*}
E\left[\frac{(1-Z_1)(1-Z_2)}{Z_1Z_2}\right]&=
\left(\frac{\alpha_4}{\alpha_2+\alpha_4}\right)\left(\frac{\alpha_3}{\alpha_1+\alpha_3}\right) \\
&+\left(\frac{\alpha_3}{\alpha_1+\alpha_3}\right)\left(\frac{\alpha_5}{\alpha_2+\alpha_4-1}\right)\\
&+\left(\frac{\alpha_4}{\alpha_2+\alpha_4}\right)\left(\frac{\alpha_5}{\alpha_1+\alpha_3-1}\right)\\
&+\left(\frac{\alpha_5}{\alpha_1+\alpha_3-1}\right)\left(\frac{\alpha_5+1}{\alpha_2+\alpha_4-1}\right) \; .
\end{align*}
Suppose 
\begin{equation} \label{eq:S}
\mathcal{S} (\tilde{\boldsymbol{z}}) = \frac{1}{n}\sum_{i=1}^{n} {\frac{(1-z_{i1})(1-z_{i2})}{z_{i1}z_{i2}}} \; .
\end{equation}
Then the sample moment at \eqref{eq:S} can be set equal to the theoretical moment.  By plugging in $\hat{a}$, $\hat{b}$, $\hat{c}$, $\hat{d}$, and ${\alpha_5}$ we have
\begin{align*}
\mathcal{S} (\tilde{\boldsymbol{z}})&= \left(\frac{{\hat{b}}-\alpha_5}{{\hat{c}}}\right) \left(\frac{{\hat{d}}-\alpha_5}{{\hat{a}}}\right)
 + \left(\frac{{\hat{d}}-\alpha_5}{\hat{a}}\right) \left(\frac{\alpha_5}{\hat{c}-1}\right)\\
&+\left(\frac{\hat{b}-\alpha_5}{\hat{c}}\right) \left(\frac{\alpha_5}{\hat{a}-1}\right)
+\left(\frac{\alpha_5(\alpha_5+1)}{(\hat{a}-1)(\hat{c}-1)}\right) \; . \label{eq:C}  
\end{align*}
This yields the quadratic equation, ${\alpha^2_5}+B{\alpha_5}+C=0$, where $B={\hat{b}}{\hat{c}}+{\hat{a}}{\hat{c}}+{\hat{a}}{\hat{d}}-{\hat{b}}-{\hat{d}}$ and
\[
C=({\hat{a}}-1)({\hat{c}}-1){\hat{b}}{\hat{d}}-{{\hat{a}}{\hat{c}}({\hat{a}}-1)({\hat{c}}-1)} \mathcal{S} (\tilde{\boldsymbol{z}}) \; .
\]
It is possible that the solution (and the estimate) for ${\alpha}_5$ can be negative, thus yielding negative values for ${\alpha}_i, i=1,2,3,4$. However, since we know that ${\alpha}_5>0$, the maximum of the larger root of the quadratic equation and $0$ will be chosen as the estimate. We apply this same principle for the estimates of ${\alpha}_i, i=1,2,3,4$. Therefore the MMLE for parameters $\alpha_i, i=1,\dots,5$ are
\begin{align}
{\hat{\alpha}_5} & = \max \left \{\displaystyle{0,\frac{-B+\sqrt{B^2-4C}}{2}}\right \} , \;\;
{\hat{\alpha}_4} = \max \left \{\displaystyle{0,\hat{b}-\hat{\alpha}_5}\right \} , \label{eq:MMLE} \\
{\hat{\alpha}_3} & = \max \left \{\displaystyle{0,\hat{d}-\hat{\alpha}_5}\right \}, \;\;
{\hat{\alpha}_2} = \max \left \{\displaystyle{0,\hat{c}-\hat{\alpha}_4}\right\} \text{, and }
{\hat{\alpha}_1} = \max \left \{\displaystyle{0,\hat{a}-\hat{\alpha}_3}\right \} \; . \notag
\end{align}

Unfortunately, $\mathcal{S} (\tilde{\boldsymbol{z}})$ is easily influenced by observed data points near zero. For example, in our simulation studies (discussed in detail later) a particular data set of size 50, denoted $\mathcal{D}$, produced the bivariate observation $(z_{43,1},z_{43,2})=(0.1089,0.0038)$. Clearly $z_{43,2}$ will severely inflate $\mathcal{S}(\tilde{\boldsymbol{z}})$, thus affecting the MMLE at \eqref{eq:MMLE}.  Further, it will affect the ABC-AR algorithm if we use $\mathcal{S}(\tilde{\boldsymbol{z}})$ as a near sufficient statistic, which we discuss in detail later.  For illustration, Table~\ref{tab:one} compares the summary statistics for $\mathcal{D}$ to those of a more typical data set, denoted $\mathcal{D}^\prime$, with no observed points near zero.  Notice that the sufficient statistics for the marginal distributions of $Z_1$ and $Z_2$ are not much affected, however there is a heavy influence on $\mathcal{S}(\tilde{\boldsymbol{z}})$.

\begin{table}[htb]
\centering
\begin{tabular}{  c  c  c  c c c c c  }
  & $  \sum \frac{\log{{z_{i1}}}}{50} $ & $ \sum \frac{\log {{z_{i2}}}}{50} $  & $ \sum \frac{\log{(1-{z_{i1}})}}{50} $ &  $ \sum \frac{\log{(1-{z_{i2}})}}{50} $ &  $\mathcal{S}(\tilde{\boldsymbol{z}})$ \\ 
\hline  
$\mathcal{D}$  & -0.81 & -1  & -1 & -0.76 & 47.77 \\

$\mathcal{D}^\prime$  & -0.76 & -0.85  & -0.76 & -0.84 & 1.67\\
\end{tabular}
\caption{Comparison of five summary statistics between dataset $\mathcal{D}$ and dataset $\mathcal{D}^\prime$.} \label{tab:one}
\end{table}

\section{Bayesian inference} \label{sec:ABC}

Suppose we have $n$ i.i.d.\ observations from the 8 parameter bivariate beta model of \cite{arnold2011flexible}.  That is, 
\[
Z_i \sim \mathcal{B}\mathcal{B}(\delta_1,\dots,\delta_8) \;.
\]
Bayesian inference requires prior distributions for ${\delta}_i > 0$, $i = 1, \dots, 8$, which we assume are independent a priori.  Our simulations consider two proper prior distributions with two hyperparameter settings.  

\begin{figure}[htb]
\centering
\includegraphics[scale=0.7]{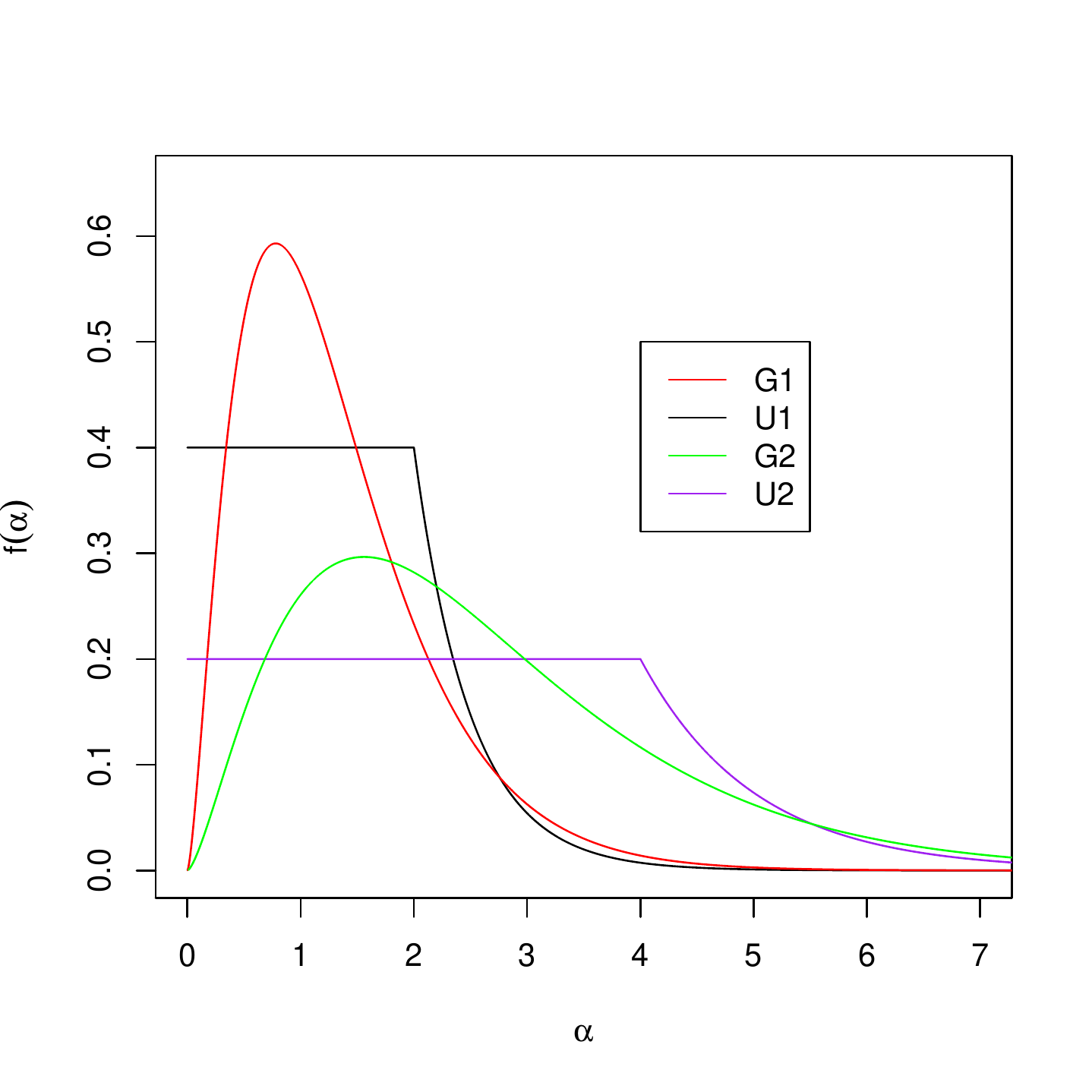}
\caption{Density curves for priors $\mathcal{G}1$ in red, $\mathcal{U}1$ in black, $\mathcal{G}2$ in green, and $\mathcal{U}2$ in purple.} \label{fig:prior}
\end{figure}

First, we consider independent modified uniform priors with support on $\R^+$ (see Figure~\ref{fig:prior}).  Specifically, for each ${\delta}_i$ the density function is
\[ 
f(\delta | \mu, p) = \left\{
  \begin{array}{l l}
    p/{\mu} & \quad \text{if } \delta {\in} (0,\mu)\\
    \frac{p}{\mu} \exp\left(\frac{-p(\delta-\mu)}{\mu(1-p)}\right) & \quad \text{if } \delta {\in} (\mu,\infty).
  \end{array} \right.\]
The motivation for the modified uniform is to reflect a lack of information of parameter values on the interval $(0,{\mu})$ where the density curve is uniform. The tail is added to cover the entire support while maintaining a proper prior.  The hyperparameter $p$ is such that $P(\delta{\in}(0,{\mu}))=p$ and $P(\delta{\in}(\mu,\infty ))=1-p$.  We denote the modified uniform as $\mathcal{U}_{p}$$(0,{\mu})$ and consider $\mathcal{U}_{0.8}$$(0,2)$ and $\mathcal{U}_{0.8}$$(0,4)$.  We denote these priors as $\mathcal{U}1$ and $\mathcal{U}2$, respectively. 

Second, we consider independent gamma priors, i.e.\ ${\delta}_i \stackrel{iid}{\sim} \Gamma({\lambda},{\beta})$, where ${\lambda}$ and ${\beta}$ are hyperparameters.  We consider $\Gamma(2.5,0.52)$ and $\Gamma(2.5,1.04)$ denoted $\mathcal{G}1$ and $\mathcal{G}2$, respectively.  To compare the gamma and modified uniform priors, the selected hyperparameters result in equal means and variances for $\mathcal{G}1$ and $\mathcal{U}1$ and for $\mathcal{G}2$ and $\mathcal{U}2$, see Figure~\ref{fig:prior}.  In the case of the reduced 5 parameter model, we consider the same independent priors for ${\alpha}_i > 0$, $i = 1, \dots, 5$.  

\subsection{Approximate Bayesian computation} 

In Bayesian analysis, the posterior density is typically known up to a normalizing constant and countless Markov chain Monte Carlo (MCMC) methods have been developed that enable sampling from the posterior \citep[see e.g.][]{brooks2011handbook}. However, these require knowledge of the likelihood function, which is unavailable for the 5 and 8 parameter bivariate beta models of \cite{arnold2011flexible}.  Bayesian inference of these models requires the likelihood-free approach of ABC, which makes sampling from the posterior (or an approximate posterior) possible. 

The fundamental idea of ABC is to generate a candidate parameter ${\boldsymbol{\theta}}^\prime$ from the prior distribution, say ${\pi(\cdot)}$, and based on this proposed parameter value, generate an auxiliary data set ${\bf y}$, i.e.\ ${\bf y} {\sim} p(\cdot \vert {\boldsymbol{\theta}^\prime)}$. If ${\bf y}$ is equal to the observed data ${\bf x}$, then ${\boldsymbol{\theta}}^\prime$ is accepted as a plausible value to have generated {\bf x}, however, if {\bf y} does not equal {\bf x}, then ${\boldsymbol{\theta}}^\prime$ is rebuffed. The accepted parameter values form an i.i.d.\ sample from the posterior distribution, say ${\pi({\boldsymbol{\theta}} \vert {\bf{x}})}$. Then if the number of accepted values, say $m$, is sufficiently large, we can form a good approximation to any Bayes estimator. 

If sufficient statistics are known, then we only require that the sufficient statistics for the auxiliary data ${\bf y}$ to be equal to the sufficient statistics of the observed data ${\bf x}$ \citep[see e.g.][]{brooks2011handbook}.  That is, we accept the proposed value if ${\bf S(y)} = {\bf S(x)}$, where ${\bf S(\cdot)}=(S_1(\cdot),\dots,S_p(\cdot))$, is the set of sufficient statistics and $p \geq dim({\boldsymbol{\theta}})$. However, in many situations sufficient statistics cannot be determined so one must rely on a set of summary statistics, thus ${\bf S(\cdot)}$ becomes a set of summary or near sufficient statistics. Furthermore, in order for ${\bf y}$ to be equal to ${\bf x}$, it is then necessary for the model to be discrete and of low dimension.
 
This is problematic for continuous models since the probability of ${\bf y}={\bf x}$ is zero. Thus, \cite{pritchard1999population} extended the above algorithm by comparing the summary statistics of ${\bf y}$ to the summary statistics of ${\bf x}$. If both sets of summary statistics are within some fixed tolerance level ${\epsilon}$, of each other, according to some distance function ${\rho}$, then the candidate parameter  ${\boldsymbol{\theta}}^\prime$ is accepted. The accepted parameter values form an i.i.d.\ sample from ${{\pi}_{\epsilon}( {\boldsymbol{\theta} \vert {\bf x}})}$ $=$ $p({\boldsymbol{\theta}} \vert \rho({\bf S(y)},{\bf S(x)})<{\epsilon} )$. The idea here is that if ${\epsilon}$ is small, then ${{\pi}_{\epsilon}( {\boldsymbol{\theta} \vert {\bf x}})}$ will provide a good approximation to ${{\pi}( {\boldsymbol{\theta} \vert {\bf x}})}$. The ABC algorithm for continuous models is described as follows\\

\begin{table}[H]
\centering
\begin{tabular}{l} 
\hline 
1. Generate ${\boldsymbol{\theta}^\prime}$ ${\sim}$ ${\pi(\boldsymbol{\cdot})}$  \\
2. Generate a data set ${\bf y}$ from the model ${p({\cdot}} \vert {\boldsymbol{\theta}^\prime)}$  \\
3. Accept ${\boldsymbol{\theta}^\prime}$ if ${\rho({\bf S(y)},{\bf S(x)})}<{\epsilon}$ otherwise discard ${\boldsymbol{\theta}^\prime}$ \\
Continue until $m$ observations have been accepted.\\
\hline
\end{tabular}
\end{table}

\noindent Thus, the outcome $({\boldsymbol{\theta}^\prime_1},\dots,{\boldsymbol{\theta}^\prime_m})$ is an i.i.d.\ sample from ${{\pi}_{\epsilon}( {\boldsymbol{\theta} \vert {\bf x}})}$. Further, the smaller the tolerance level, the greater the computational cost, as will be evident in our simulation study.

A deficiency of the ABC-AR algorithm is that while the accepted parameter values are independent it provides low acceptance rates \citep[see e.g.][]{beaumont2009adaptive}.  To overcome this problem, \cite{marjoram2003markov} proposed implementing the Metropolis Hastings algorithm into the ABC algorithm (ABC-MH). However, while acceptance rates can be improved this method often leads to highly correlated draws.  Our simulations consider a random walk ABC-MH algorithm with a Normal proposal described as follows\\

\begin{table}[H]
\centering
\begin{tabular} {l}
\hline 
1. Initialize ${\boldsymbol{\theta}^{(1)}}$, $m=1$ \\
2. Generate independent candidate parameters ${\theta_i^\prime}$ ${\sim}$ $N(\theta_i^{(m)},\sigma_i^2 )$\\
\quad for $i=1,\dots,dim({\boldsymbol{\theta}})$\\
3. Generate a data set ${\bf y}$ from the model ${p({\cdot}} \vert {\boldsymbol{\theta}^\prime)}$  \\
4. Set ${\boldsymbol{\theta}^{(m+1)}}={\boldsymbol{\theta}^\prime}$ with probability\\
\quad $h=\left\{ 1 , \displaystyle \frac{\pi({\boldsymbol{\theta}^\prime})}{ \pi({\boldsymbol{\theta}^{(m)}}) } 
\mathbb{I} \left( \rho({\bf S(y)},{\bf S(x)})< \epsilon \right)
\right\}$,\\
\quad otherwise set ${\boldsymbol{\theta}^{(m+1)}}={\boldsymbol{\theta}^{(m)}}$\\
5. Set $m = m+1$\\
Continue until $m$ reaches desired number of iterations.\\
\hline 
\end{tabular}
\end{table}

To improve upon the inefficiency of ABC-MH, there have been a large number of proposals that incorporate sequential Monte Carlo (SMC) techniques. \cite{sisson2007sequential} was one of the first to make use of SMC methodology, proposing coupling SMC with partial rejection control and a biased approximation of the posterior distribution. \cite {beaumont2009adaptive} further proposed utilizing population Monte Carlo methods of \cite{cappe2004population}.  Similarly, \cite{toni2009approximate} proposed an algorithm derived from the framework of sequential importance sampling.  % It should be noted that  \cite{sisson2007sequentialcorrection} offered a correction which alleviated the problem. 
\cite{peters2012sequential} embed the partial rejection control mechanism of \cite{liumonte} which incorporates a mutation and correction step within the standard SMC sampler algorithm. This incorporation of a mutation kernel reduces the variability of the importance weights when compared to more standard SMC algorithms. 

Another attempt to improve ABC-AR is to incorporate regression methodology. The pioneer of this approach was \cite{beaumont2002approximate} who assumed that the conditional density can be described by a regression model. The idea was to weight the parameters by comparing the auxiliary summary statistics with the observed summary statistics.  An interested reader is directed to \cite{blum2010non} and \cite{leuenberger2010bayesian} for extensions of this approach.
Unfortunately, these methods focus on univariate settings though some authors comment that an extension using multivariate regression is straightforward. 

Our work considers both the ABC-AR and ABC-MH algorithms under various parameter settings, priors, sample sizes, and tolerance levels.  The use of SMC and regression methodology is a direction of future research.

\subsection{ABC for the bivariate beta model}

Given the joint likelihood is unavailable in closed form, sufficient statistics cannot be determined.  Thus, we are forced to choose informative summary or near sufficient statistics.  First, consider the 5 parameter model where we will use 5 summary statistics.  Since the marginals of $Z_1$ and $Z_2$ are distributed as beta random variables, we choose the corresponding univariate sufficient statistics, i.e. $S_1(\tilde{\boldsymbol{z}}) = \frac{1}{n}\sum \log {{z_{i1}}}$, $S_2(\tilde{\boldsymbol{z}}) = \frac{1}{n}\sum \log {{z_{i2}}}$, $S_3(\tilde{\boldsymbol{z}}) = \frac{1}{n}\sum \log {(1-{z_{i1}})}$, and $S_4(\tilde{\boldsymbol{z}}) = \frac{1}{n}\sum \log {(1-{z_{i2}})}$.

For our fifth summary statistic, we first considered $\mathcal{S} (\tilde{\boldsymbol{z}})$ at \eqref{eq:S} used in MMLE.  As we have illustrated in Table~\ref{tab:one}, $\mathcal{S} (\tilde{\boldsymbol{z}})$ is influenced by small observed values. The implications in a preliminary simulation study were twofold. First, a small portion of simulated datasets contained severely inflated values of $\mathcal{S} (\tilde{\boldsymbol{z}})$, one of which is illustrated in Table~\ref{tab:one}. Thus, there would be difficulty generating an auxiliary dataset where $\mathcal{S} (\tilde{\boldsymbol{z}})$ is close to that of the observed data set resulting in extremely low acceptance rates. Second, using $\mathcal{S} (\tilde{\boldsymbol{z}})$ as a summary statistic inflated the observed bias and MSE in repeated simulations.

Given the problems with $\mathcal{S} (\tilde{\boldsymbol{z}})$ as a summary statistic, we require an alternative to capture the correlation that offers more stability for any observed values.  To this end, we used the Pearson correlation between $\vec{\boldsymbol{z}}_1$ and $\vec{\boldsymbol{z}}_2$, that is
\[
S_5(\tilde{\boldsymbol{z}}) = \frac{\sum(z_{i1}-\bar{z}_{\cdot1})(z_{i2}-\bar{z}_{\cdot2})}{\sqrt{\sum(z_{i1}-\bar{z}_{\cdot1})^2\sum(z_{i2}-\bar{z}_{\cdot2})^2}}.
\]
Thus, $S = \left( S_1(\tilde{\boldsymbol{z}}), S_2(\tilde{\boldsymbol{z}}), S_3(\tilde{\boldsymbol{z}}), S_4(\tilde{\boldsymbol{z}}), S_5(\tilde{\boldsymbol{z}}) \right)$ is the vector of summary statistics used for ABC in the 5 parameter model.  The use of $S_5(\tilde{\boldsymbol{z}})$ instead of $\mathcal{S} (\tilde{\boldsymbol{z}})$ vastly improved the acceptance rates, bias and MSE.  

Finally, we considered the distance function 
\[
{\rho(S(\tilde{\boldsymbol{z}}),S(\tilde{\boldsymbol{y}}))} = \sum^5_{i=1} \left| S_i(\tilde{\boldsymbol{z}})-S_i(\tilde{\boldsymbol{y}}) \right|
\]
where $\tilde{\boldsymbol{y}}$ denotes the auxiliary dataset.  Preliminary simulations showed the 5 summary statistics had approximately equal variability for a variety of distance cutoff values.  Hence, there is no need to consider weights or a scale adjustment in the distance function.

Next, consider the 8 parameter model where we will use 8 summary statistics.  We begin by including the 5 summary statistics from the smaller model.  In order to capture additional dependency between $Z_1$ and $Z_2$, we added the Spearman rank correlation and Kendall correlation, that is
\[
S_6(\tilde{\boldsymbol{z}}) = 1-\frac{6\sum{d^2_i}}{n(n^2-1)} \text{ where } {d_i=z_{i1}-z_{i2}} \text{ and }
\]
\[
 S_7(\tilde{\boldsymbol{z}}) =\frac{(\text{number of concordant pairs})-(\text{number of discordant pairs})}{\frac{1}{2}n(n-1)} \; .
\]
Finally, we consider $S_8(\tilde{\boldsymbol{z}}) = \frac{1}{n}\sum \sqrt {z_{i1}z_{i2}}$ as our eighth summary statistic. Our distance function is ${\rho(S(\tilde{\boldsymbol{z}}),S(\tilde{\boldsymbol{y}}))} = \sum^8_{i=1} \left| S_i(\tilde{\boldsymbol{z}})-S_i(\tilde{\boldsymbol{y}}) \right|$.  Again, preliminary simulations showed the 8 summary statistics were approximately equal in terms of scale and variability.

\section{Simulation study} \label{sec:sim}

This section investigates parameter estimation for the 5 and 8 parameter bivariate beta models through a variety of simulations.  In total, we considered 96 and 32 settings for the 5 and 8 parameter models, respectively.  In each setting, we independently repeat the simulation for 200 simulated datasets to evaluate the resulting parameter estimates.  Specifically, estimated posterior means obtained via ABC are compared to the true values used to simulate the data.  For the 5 parameter model, Bayesian estimates are also compared to those obtained via MMLE.

Given the breadth of this simulation study, we only present results using the ABC-AR algorithm.  We note that the ABC-MH algorithm (results not shown) sometimes resulted in highly correlated samples and low acceptance rates caused by poor starting values or proposal variances.  Practitioners, with a single data set, may find the cost of tuning an ABC-MH algorithm or incorporating SMC techniques a worthwhile investment.  Section~\ref{sec:Example} considers such an example comparing results from the ABC-AR to ABC-MH.

Overall, as $\epsilon$ decreases we observe decreases in bias and MSE relative to the true values.  However, this improvement requires additional computational effort.  The choice of prior also has a significant impact on both bias and computational time.  In short, priors centered close to the observed data result in less bias and greater computational efficiency of the ABC algorithm.  In our settings, gamma priors improve computational efficiency slightly and lead to smaller MSEs.  For this reason, we suggest a gamma prior in conjunction with a ``small'' $\epsilon$.  

There are a number of other potential priors including uniform, triangle, Epanechnikov or Gaussian (truncated on $\R^+$) kernels.  In applications, these will impact acceptance rates (positively and negatively) for an ABC-AR algorithm, which is a direction of future research.  The following sections outline specific results for the 5 and 8 parameter model.  

\subsection{5 parameter model} \label{sec:sim 5}

\begin{figure}
\centering
\subfloat[$A_1$, $r$=-0.2384]{\includegraphics[scale=0.45]{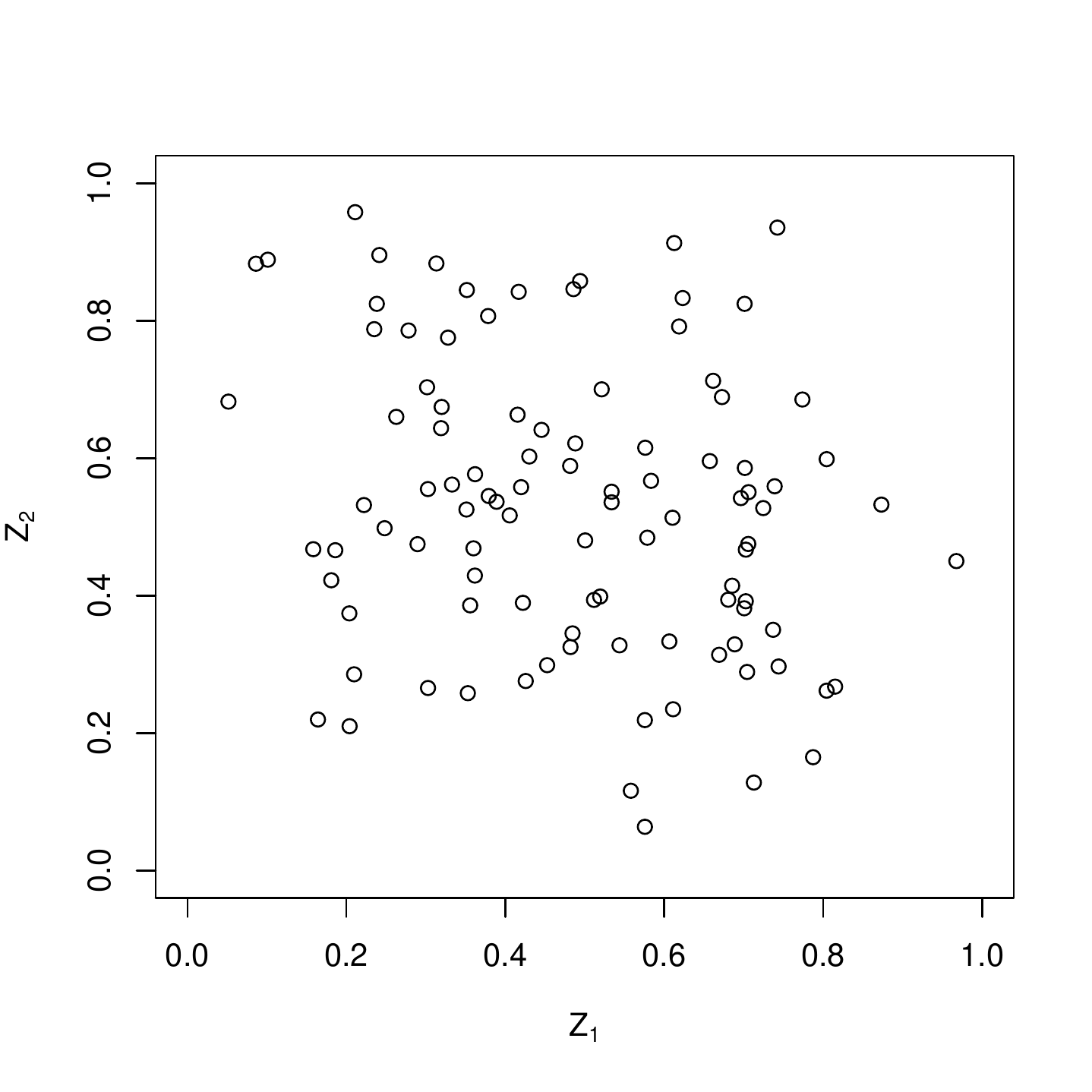}\label{fig:oneA}}
\subfloat[$A_2$, $r$=-0.2859]{\includegraphics[scale=0.45]{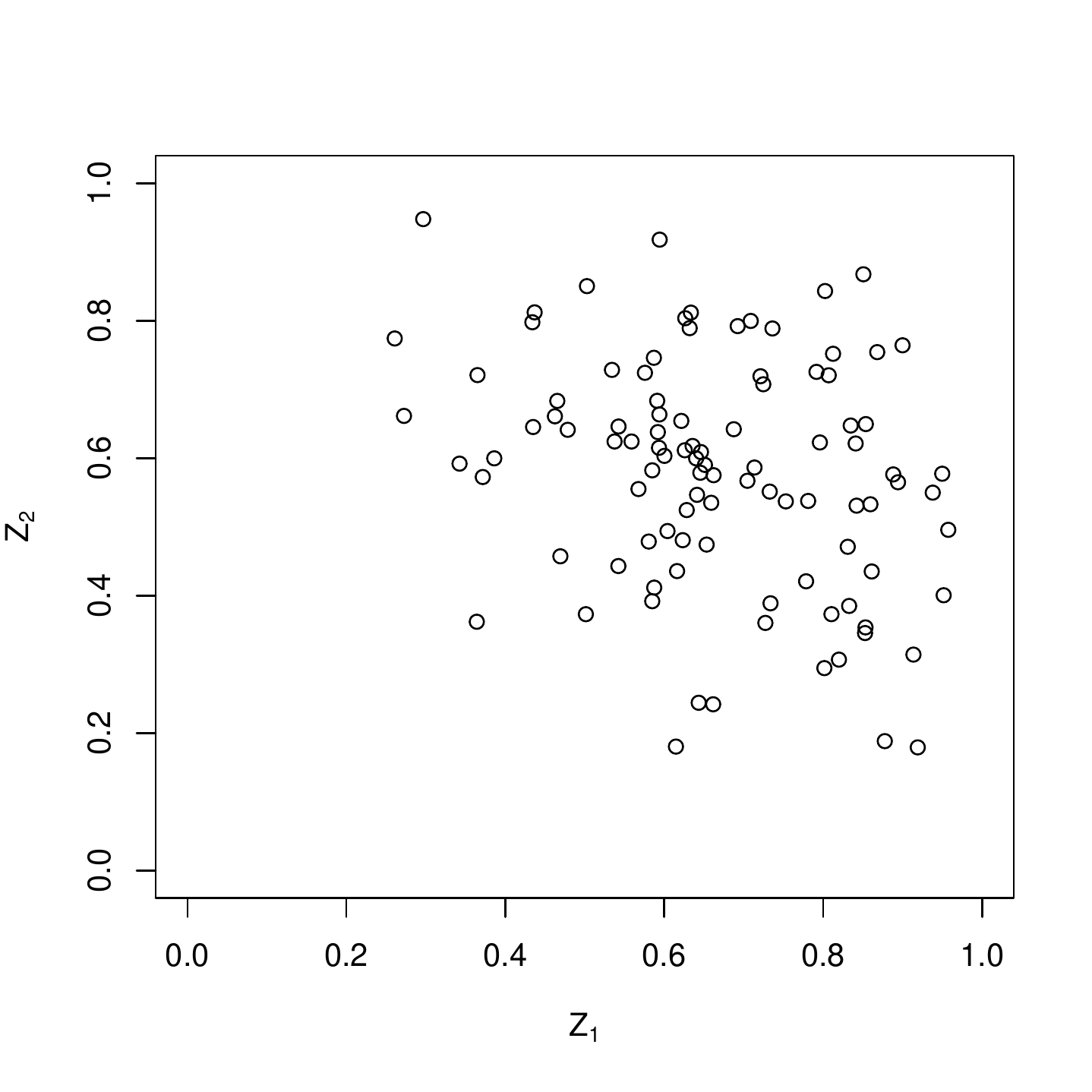}\label{fig:oneB}}\\
\subfloat[$A_3$, $r$=-0.7068]{\includegraphics[scale=0.45]{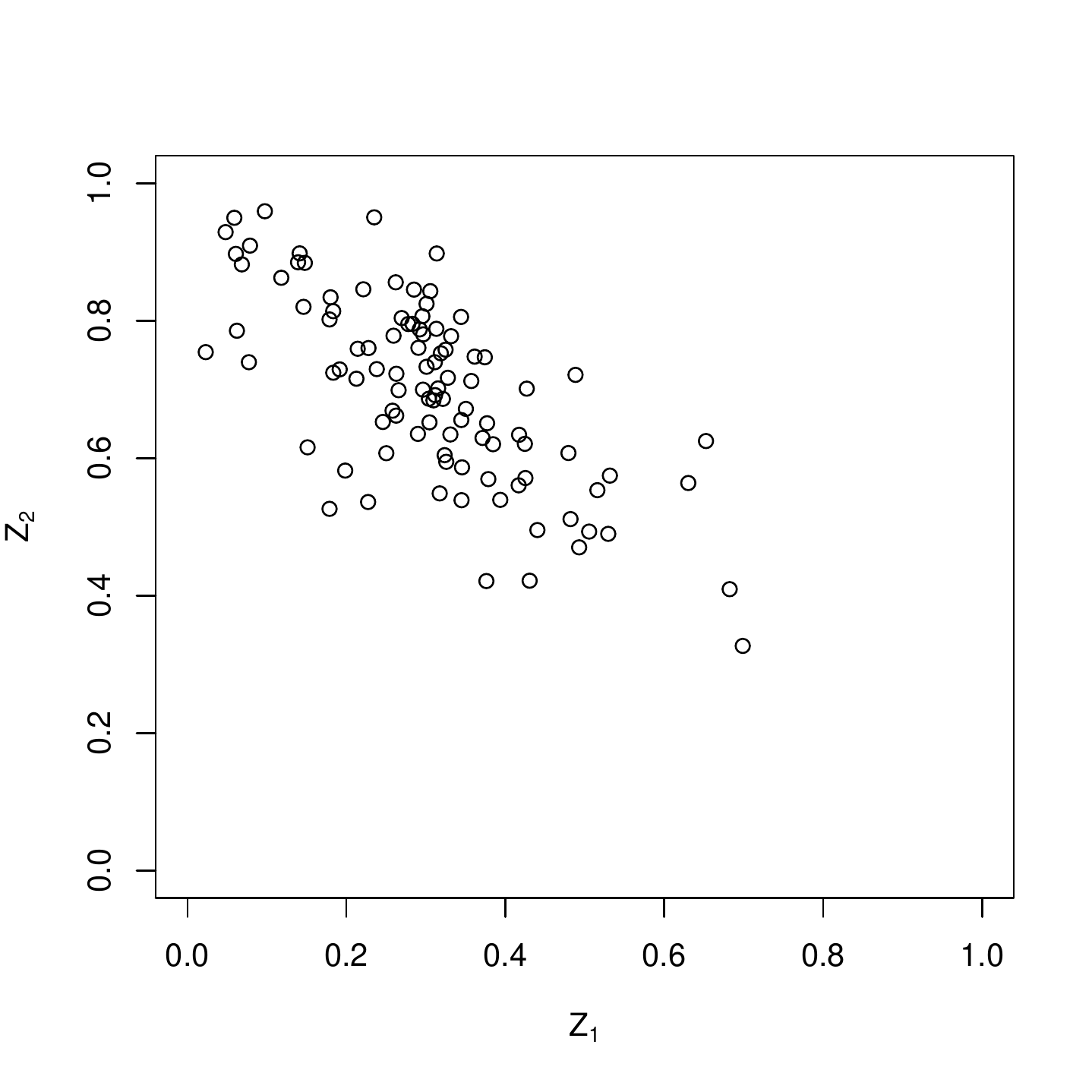}\label{fig:oneD}}
\caption{Scatterplots of $Z_1$ and $Z_2$ for $n=100$ with the estimated correlation.} \label{fig:scatter}
\end{figure}

For the 5 parameter model, we consider three parameter settings, $A_1 = (1,1,1,1,1)^\prime$, $A_2 = (3,2.5,2,1.5,1)^\prime$, and $A_3 = (1,1,2,6,1)^\prime$. The first two settings were used by \cite{arnold2011flexible}.  Figure~\ref{fig:scatter} shows a scatter plot for a single data set for each parameter setting.  Note that each of these has a negative correlation. 

Within each parameter setting, we consider every combination of prior in $\left \{ \mathcal{G}1, \mathcal{G}2, \mathcal{U}1, \mathcal{U}2 \right \}$, $n \in \left \{50,100 \right \}$, and $\epsilon \in \left \{0.2, 0.4, 0.6, 0.8 \right \}$.  Each setting was repeated across 200 datasets with inferences based on 1000 ABC-AR acceptances (or 15e6 proposals, whichever came first). %Estimated marginal posterior means were used to estimate $\alpha_i, i=1,\dots,5$ as well as MMLE, which we compare to each other and the true values using bias and MSE.

Table~\ref{tab:5 A1} shows the results for $A_1$ when $n=100$.  As $\epsilon$ decreases, bias and MSE decrease while the number of proposals increases.  The choice of prior also impacts the number of proposals required, with priors centered near the true values being more efficient.  For example, when ${\epsilon}=0.2$, the $\mathcal{G}1$ and $\mathcal{G}2$ priors required approximately 9e5 and 1e7 proposals (on average), respectively.  

\begin{figure}[p]
\centering
\subfloat[${\hat{\alpha}_1}$]{\includegraphics[trim=0 35 50 35,clip, scale=0.45]{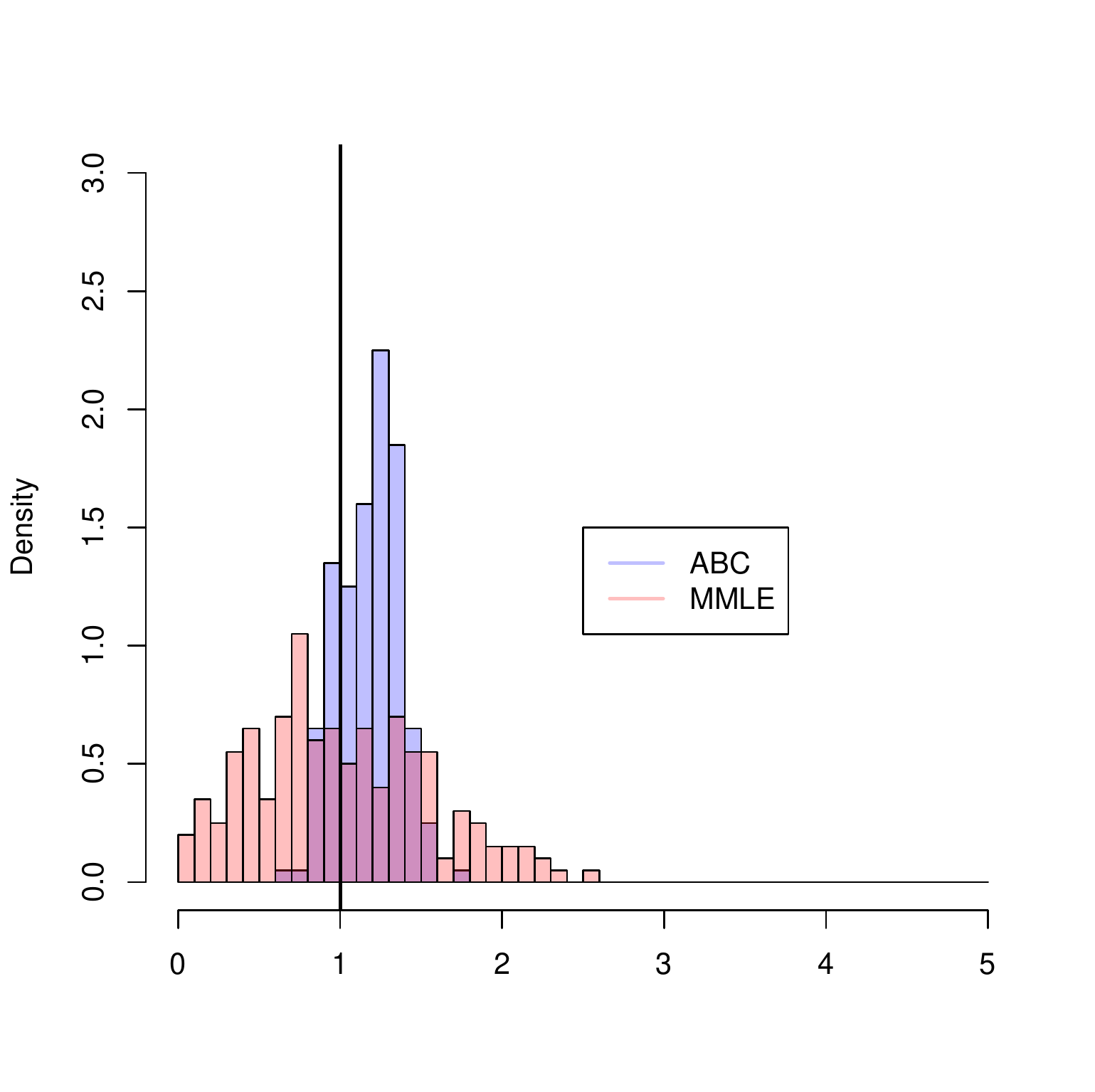}\label{fig:oneA}}
\subfloat[${\hat{\alpha}_2}$]{\includegraphics[trim=0 35 50 35,clip, scale=0.45]{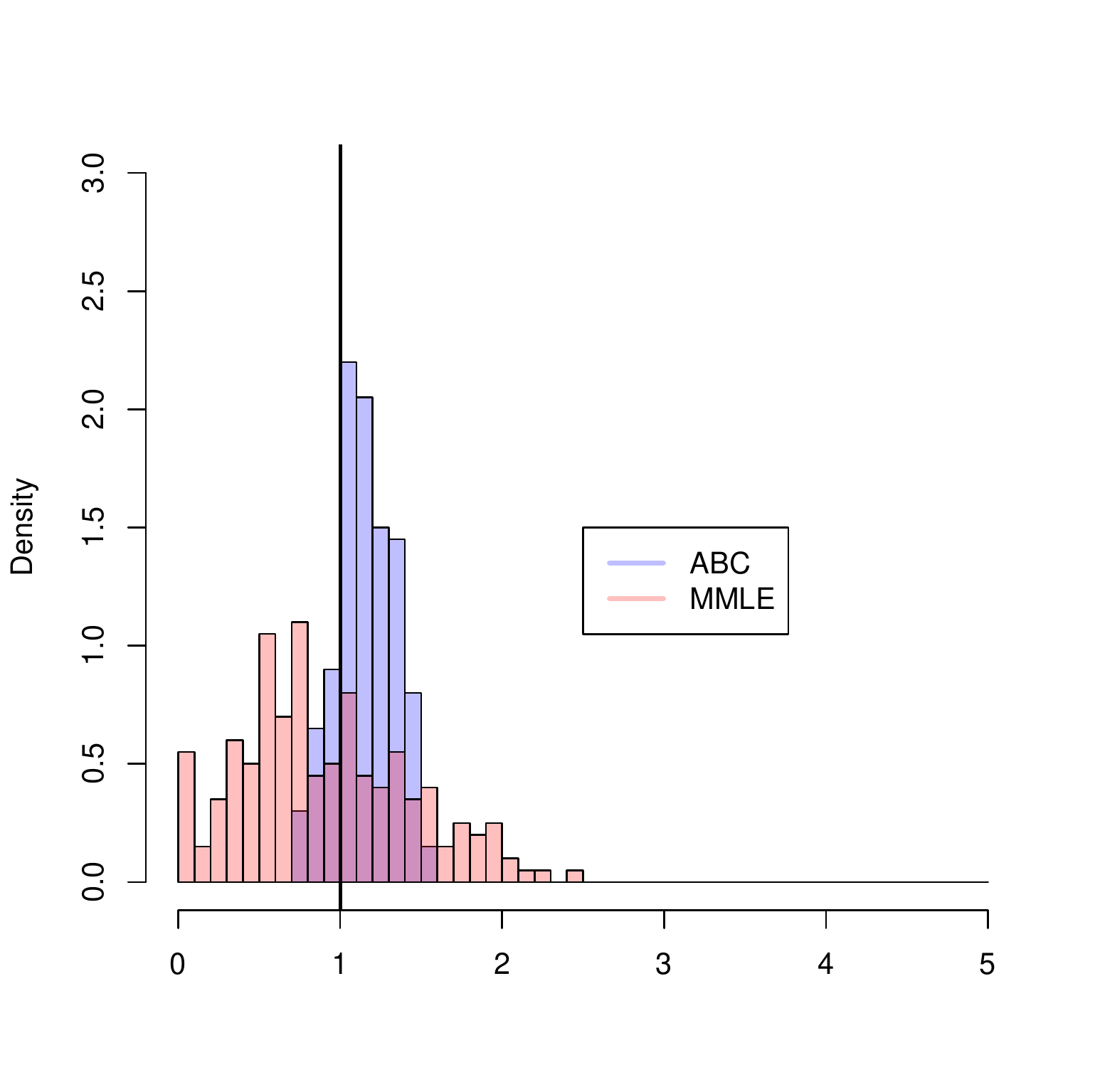}\label{fig:oneB}}\\
\subfloat[${\hat{\alpha}_3}$]{\includegraphics[trim=0 35 50 35,clip, scale=0.45]{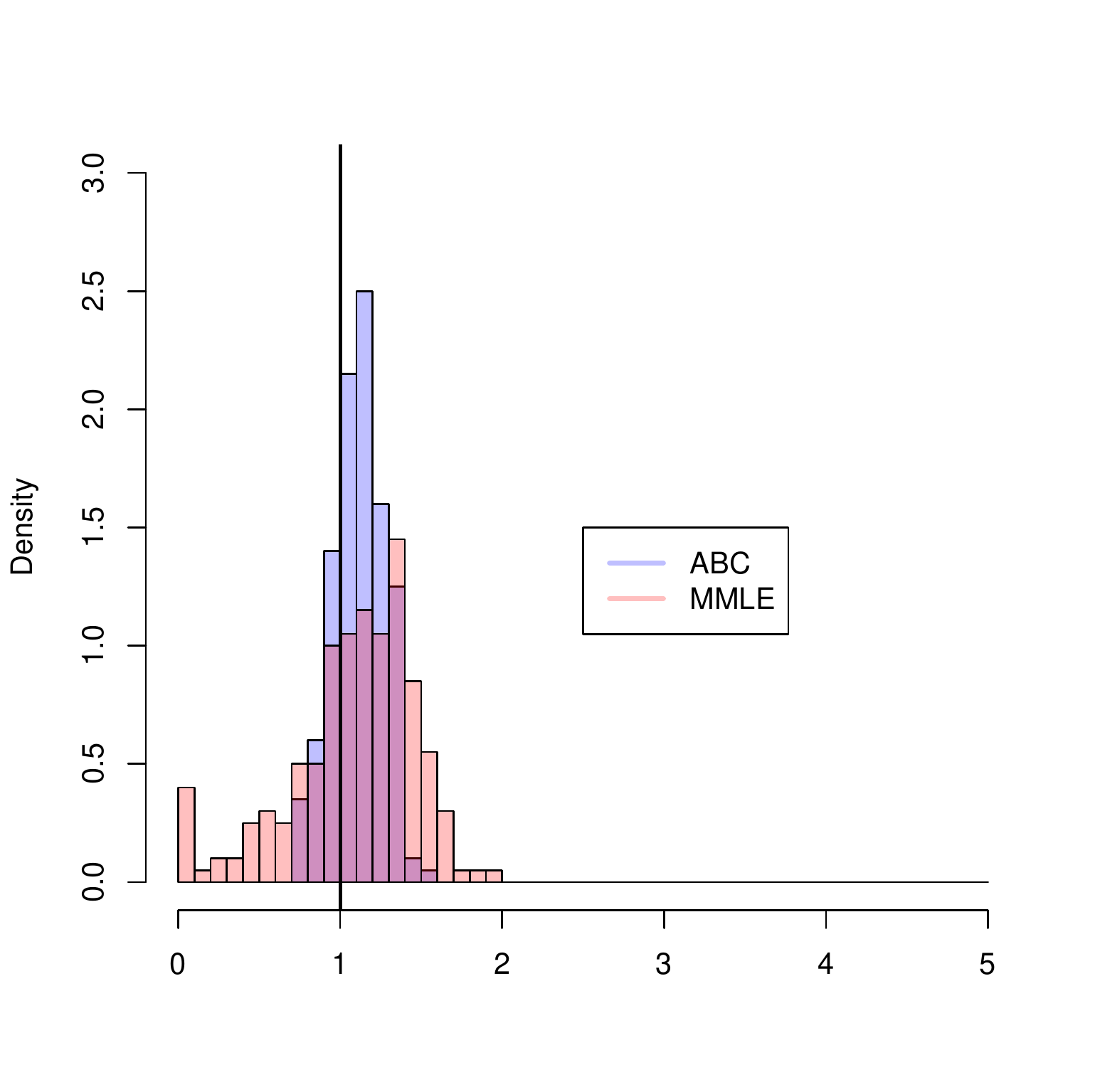}\label{fig:oneD}}
\subfloat[${\hat{\alpha}_4}$]{\includegraphics[trim=0 35 50 35,clip, scale=0.45]{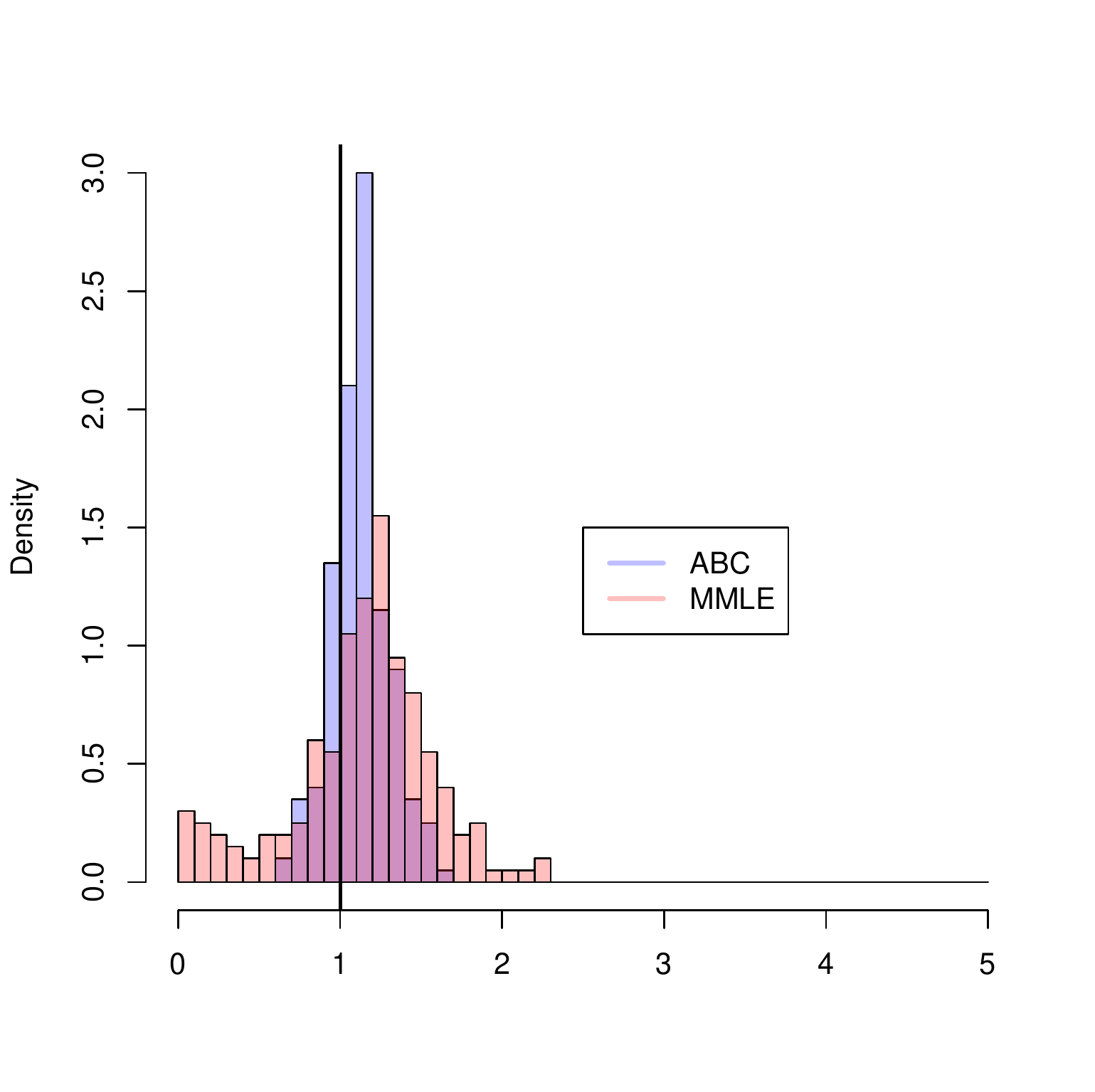}\label{fig:oneE}}\\
\subfloat[${\hat{\alpha}_5}$]{\includegraphics[trim=0 35 50 35,clip, scale=0.45]{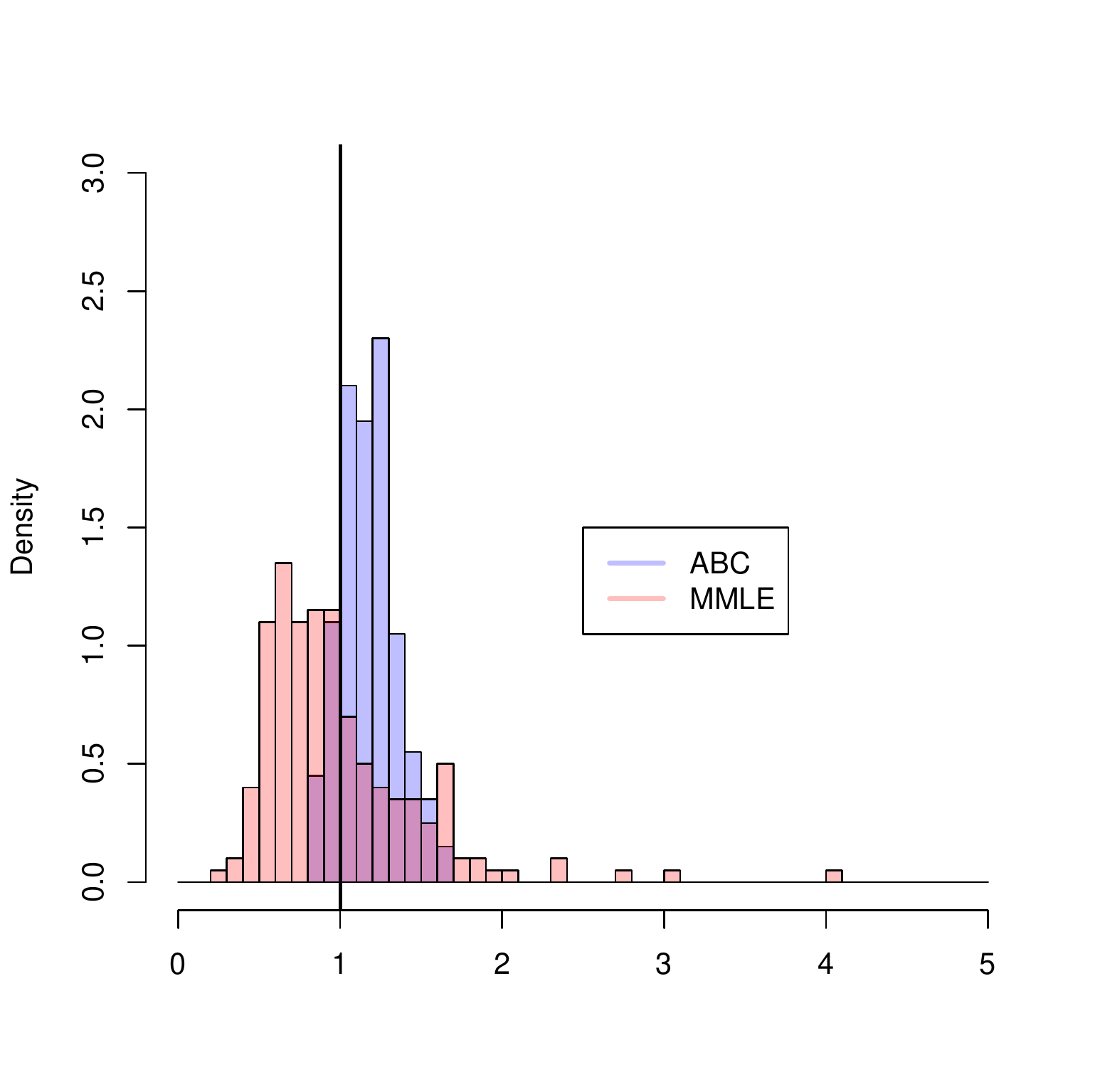}\label{fig:oneF}}
\caption{Histograms comparing ABC-AR with the $\mathcal{G}1$ prior to MMLE under $A_1$, $n=100$, and $\epsilon=0.2$.} \label{fig:hist}
\end{figure}

As expected, the Bayesian approach introduces bias not present with MMLE.  However, even in the presence of bias the overall precision of estimation improves using the Bayesian approach.  To illustrate this, Figure~\ref{fig:hist} shows histograms of estimators based on MMLE and ABC for the 200 simulated datasets.  The histograms were generated under the $\mathcal{G}1$ prior, $n=100$ and ${\epsilon}=0.2$ and the black vertical line represents the true parameter value.  Here, we can clearly see the bias from a Bayesian approach and the reduction in variability.  One reason for this reduction is elimination of $\mathcal{S} (\tilde{\boldsymbol{z}})$ when using ABC.  Overall, estimating $(\alpha_1,\dots,\alpha_5)$ with posterior expectations decreased MSE relative to MMLE for every prior considered.  Similar conclusions were observed for the case when $n=50$ (results not shown).

Table~\ref{tab:5 A2} shows the results for $A_2$ when $n=100$.  As with the previous setting, bias and MSE decrease as $\epsilon$ decreases at the expense of computational time.  The Bayesian approach also introduces bias for all priors.  For $\mathcal{U}1$ and $\mathcal{G}1$, we observe smaller MSE relative to MMLE, however for $\mathcal{U}2$ and $\mathcal{G}2$ the MSE results are similar to MMLE.  Finally, we observe the overall computational costs for $\mathcal{U}2$ and $\mathcal{G}2$ under $A_2$ are lower than that of $A_1$.  Similar conclusions were observed when $n=50$.  

Results for $A_3$ show similar behavior as the other settings (results not shown), but are slightly more muted.  The challenge of this setting is having ${\alpha_4}=6$ in the model.  In this case, $\mathcal{G}2$ and $\mathcal{U}2$ result in lower MSE and simulation effort since they are more likely to propose values near ${\alpha_4}$.

\subsection{8 parameter model} \label{sec:sim 8}

For the 8 parameter model, we considered parameter settings $A_4 = (2,1,1,2,4,6,2,1)^\prime$ and $A_5 = (3.5,2,1.5,4,1,2.5,3,4.5)^\prime$ and set $n=100$.  Here we consider every combination of prior in $\left \{ \mathcal{G}1, \mathcal{G}2, \mathcal{U}1, \mathcal{U}2 \right \}$ and $\epsilon \in \left \{0.2, 0.4, 0.6, 0.8 \right \}$.  Each setting was repeated across 200 datasets with inferences based 1000 ABC-AR acceptances (or 15e6 proposals, whichever came first).  Again estimated posterior means were used to estimate $\delta_i, i=1,\dots,8$, which were compared to the true values using bias and MSE.  The results were not compared to an existing approach since we know of no other applicable method.

Table~\ref{tab:8 A4} displays the results for the $A_4$ setting.  Again, bias and MSE decrease as ${\epsilon}$ decreases.  Note there is a significant amount of bias with large MSEs for ${\alpha_5}$ and ${\alpha_6}$ under the $\mathcal{G}1$ and $\mathcal{U}1$ priors since they tend to propose values far from the truth.  However, the estimation improves under the $\mathcal{G}2$ and $\mathcal{U}2$ priors.  We observed similar behaviour for the $A_5$ setting.

\section{Bacon and eggs} \label{sec:Example}

In this section, we apply our bivariate beta model to an example previously analyzed by \cite{danaher2005bacon}. The objective of the study was to observe the behavior of households and their grocery store habits.  In particular, we study the probabilities and correlation of purchasing bacon and eggs on a single shopping trip. In the study, a sample of 548 independent households were taken and details of what the household purchased at the market were recorded over 4 consecutive trips. For each trip, it was recorded whether or not the household purchased bacon or eggs or both, see Table~\ref{tab:data}.  We will refer to Table~\ref{tab:data} as $\mathcal{T}$, which is a 5x5 matrix of observations. 

\begin{table}[htb]
\centering
\begin{tabular}{c|ccccc|c}

\multicolumn {7}{c} {Eggs} \\
 Bacon       & 0    & 1     & 2    & 3     & 4    & Total \\ \hline  
0  & 254 & 115 & 42 & 13 & 6 & 430 \\ 
1  & 34 & 29 & 16 & 6 & 1 & 86 \\ 
2  & 8 & 8 & 3 & 3 & 1 & 23 \\ 
3  & 0 & 0 & 4 & 1 & 1 & 6 \\ 
4  & 1 & 1 & 1 & 0 & 0 & 3 \\ 
\hline
Total  & 297 & 153 & 66 & 23 & 9 & 548 \\ 
\end{tabular}
\caption{Bivariate binomial counts describing bacon and egg purchases.} \label{tab:data}
\end{table}

Let $X_{kb}$ and $X_{ke}$ represent the number of times the $kth$ customer purchased bacon and eggs over the course of the 4 trips, respectively. Clearly, $X_{kb}$ and $X_{ke}$ are correlated, and so \cite{danaher2005bacon} proposed a bivariate beta binomial model to capture the over dispersion and correlation.  Let $p_{kb}$ and $p_{ke}$ denote the probability of household $k$ purchasing bacon and eggs, respectively. In this model, $(p_{kb},p_{ke})$ is a bivariate random vector, where the requirement is that it follow some bivariate joint density, where the marginals are beta distributed. Thus,  $(X_{kb},X_{ke}) \vert (p_{kb},p_{ke})$ $\sim$ $\mathcal{B}iv\mathcal{B}in(4,p_{kb},p_{ke})$ where $\mathcal{B}iv\mathcal{B}in$ denotes a bivariate binomial distribution. Furthermore, $X_{kb}$ and $X_{ke}$ are conditionally independent given $(p_{kb},p_{ke})$, i.e. $X_{kb} \vert p_{b} {\sim}Bin(4,p_{kb})$ and $X_{ke} \vert p_{e} {\sim}Bin(4,p_{ke})$.  The unconditional correlation between $X_{kb}$ and $X_{ke}$ is introduced through the bivariate distribution of $(p_{kb},p_{ke})$.

We propose use of the 5 parameter bivariate beta model proposed by \cite{arnold2011flexible}, that is 
\[
(p_{kb},p_{ke}) {\sim} \mathcal{B B} ( \alpha_1, \dots, \alpha_5 ) \; .
\]
Furthermore, we consider gamma priors for $\alpha_i$ $i = 1, \dots, 5$ and use an empirical Bayes approach to select hyperparameters.  Specifically, we compute marginal MLEs under the beta binomial model and use this to help guide our hyperparameter selection.  Using the beta binomial distribution family function within the \verb=VGAM= package in \verb=R=, we have the following MLEs, $\tilde{\alpha}_{b}=0.3571$, $\tilde{\beta}_{b}=4.4552$, $\tilde{\alpha}_{e}=0.8592$, and $\tilde{\beta}_{e}=3.9593$. Linking our bivariate beta parameters to these estimates we have
\begin{align}
{\widetilde{\alpha_1+\alpha_3}}={\tilde{\alpha}_{b}}=0.3571 & \hspace{1cm}
{\widetilde{\alpha_4+\alpha_5}}={\tilde{\beta}_{b}}=4.4552 \notag \\
{\widetilde{\alpha_2+\alpha_4}}={\tilde{\alpha}_{e}}=0.8592 & \hspace{1cm}
{\widetilde{\alpha_3+\alpha_5}}={\tilde{\beta}_{e}}=3.9593  \; . \label{constraints1}
\end{align}

We choose values of $\tilde{\alpha}_i, i=1,\dots,5$ such that $\alpha_i$ is centered around $\tilde{\alpha}_i$ a priori with variance of 1, i.e.\ ${\alpha}_i$ ${\sim}$ $\Gamma(\tilde{\alpha}_i^2,1/\tilde{\alpha}_i), i=1,\dots,5$. Furthermore, since we believe there should be a moderate and positive correlation between the purchase of bacon and eggs, we choose $\tilde{\alpha}_i, i=1,\dots,5$ such that the Monte Carlo correlation estimate under $\mathcal{B}\mathcal{B}(\tilde{\alpha}_1,\dots,\tilde{\alpha}_5)$ is close to 0.30.  In short, we choose values close to \eqref{constraints1} with a correlation near 0.30.  The exact prior means are, $\tilde{\alpha}_1=1.6182$, $\tilde{\alpha}_2=1.9932$, $\tilde{\alpha}_3=0.1684$, $\tilde{\alpha}_4=0.1702$, and $\tilde{\alpha}_5=3.1234$, where the Monte Carlo estimate of the correlation is 0.3004. As we will see, this choice allows for exploration of the parameter space with reasonable computational effort.

Thus, the Bayesian hierarchical model contains the following stages
\begin{align}
(X_{kb},X_{ke}) \vert (p_{kb},p_{ke}) & \sim \mathcal{B}iv\mathcal{B}in(4,p_{kb},p_{ke}) \text{ for } k = 1, \dots, 548, \notag \\
(p_{kb},p_{ke}) & \sim \mathcal{B B} ( \alpha_1, \dots, \alpha_5 ) \text{ for } k = 1, \dots, 548, \label{posterior}\\
{\alpha}_i & \sim \Gamma(\tilde{\alpha}_i^2,1/\tilde{\alpha}_i)  \text{ for } i=1, \dots, 5. \notag
\end{align}

\cite{danaher2005bacon} proposed using the bivariate beta model from \cite{sarmanov1966} which can be described as $g(p_{b},p_{e})=f_b(p_{b})f_e(p_{e})[1+\omega\phi_b(p_{b})\phi_e(p_{e})]$, where $\phi_b(p_{b})$ is a bounded non-constant ``mixing'' function such that $\int{\phi_b(l)f_b(l)dl}$=0 (similar for ``eggs'').  The parameter $\omega$ determines the correlation between $p_{b}$ and $p_{e}$ and must satisfy the condition $1+\omega\phi_b(p_{b})\phi_e(p_{e})>0$ for all $p_{b}$ and $p_{e}$ to be a valid joint density function. Furthermore, the marginals are beta distributed, i.e.\ $p_{b} {\sim} Beta(\alpha_b,\beta_b)$ and $p_{e} {\sim} Beta(\alpha_e,\beta_e)$.  Letting $\phi_b(p_{b})=p_{b}-\mu_b$, where $\mu_b=E(p_b)=\frac{\alpha_b}{\alpha_b+\alpha_b}$, and similarly for ``eggs,'' yields a closed form likelihood enabling estimation via maximum likelihood. 

\subsection{Sampling algorithms}

Given the prior density for $(p_{kb},p_{ke})$ is unavailable in closed form, we consider the ABC-AR and ABC-MH algorithms to obtain posterior samples from the model at \eqref{posterior}.  Since our data is discrete, it is possible to simulate $\mathcal{T}$ exactly.  However, since the probability of this event is very small, we will accept candidate parameters when the auxiliary table $\mathcal{T}^\prime$ is close to the observed table $\mathcal{T}$ . Here, we must define what ``close'' means, i.e.\ what does it mean for $\mathcal{T}^\prime$ $\approx$ $\mathcal{T}$?  To this end, we consider the absolute difference between the cells of $\mathcal{T}^\prime$ and $\mathcal{T}$. In other words, let $\mathcal{T}=\{ a_{lj} \}$ and $\mathcal{T}^\prime=\{ b_{lj} \}$ and the distance function be $\rho = \sum^5_{l=1} \sum^5_{j=1} \left| a_{lj} - b_{lj} \right|$.  

For the ABC-AR algorithm, the simulation was run until 500 acceptances with $\epsilon=100$ and is described as follows\\

\begin{table}[H]
\centering
\begin{tabular}{l} 
\hline 
1. Generate ${\alpha}_i^\prime \sim \Gamma(\tilde{\alpha}_i^2,1/\tilde{\alpha}_i)$ for $i=1, \dots, 5$ \\
2. Generate $(p_{kb},p_{ke}) \sim \mathcal{B B} ( \alpha_1^\prime, \dots, \alpha_5^\prime )$ for $k=1, \dots, 548$\\
3. Generate $(Y_{kb},Y_{ke}) \vert (p_{kb},p_{ke}) \sim \mathcal{B}iv\mathcal{B}in(4,p_{kb},p_{ke})$ for $k = 1, \dots, 548$\\
4. Generate an auxiliary table $\mathcal{T}^\prime \vert (Y_{1b},Y_{1e}), \dots, (Y_{548b},Y_{548e})$\\
5. Accept $(\alpha_1^\prime, \dots, \alpha_5^\prime)$ if $\rho = \sum^5_{l=1} \sum^5_{j=1} \left| a_{lj} - b_{lj} \right|<\epsilon$ otherwise discard \\
Continue until $500$ observations have been accepted.\\
\hline 
\end{tabular}
\end{table}

For the ABC-MH algorithm, the simulation was run for $2e6$ iterations with $\epsilon=100$. We considered a Normal proposal random walk with component standard deviations 0.10, 0.10, 0.001, 0.001, and 0.2 for $i=1,\dots,5$, respectively, and is described as follows\\

\begin{table}[H]
\centering
\begin{tabular} {l}
\hline 
1. Initialize $(\alpha_1^{(1)}, \dots, \alpha_5^{(1)})$, $m=1$ \\
2. Generate candidate parameters ${\alpha_i^\prime}$ ${\sim}$ $N(\alpha_i^{(m)},\sigma_i^2 )$ for $i=1,\dots,5$\\
3. Generate $(p_{kb},p_{ke}) \sim \mathcal{B B} ( \alpha_1^\prime, \dots, \alpha_5^\prime )$ for $k=1, \dots, 548$\\
4. Generate $(Y_{kb},Y_{ke}) \vert (p_{kb},p_{ke}) \sim \mathcal{B}iv\mathcal{B}in(4,p_{kb},p_{ke})$ for $k = 1, \dots, 548$\\
5. Generate an auxiliary table $\mathcal{T}^\prime \vert (Y_{1b},Y_{1e}), \dots, (Y_{548b},Y_{548e})$\\
6. Set $(\alpha_1^{(m+1)}, \dots, \alpha_5^{(m+1)}) = (\alpha_1^\prime, \dots, \alpha_5^\prime)$ with probability\\
\quad $h=\left \{\displaystyle{1,\frac{ \pi(\alpha_1^\prime, \dots, \alpha_5^\prime)}{\pi(\alpha_1^{(m)}, \dots, \alpha_5^{(m)})} \mathbb{I}{(\rho = \sum^5_{l=1} \sum^5_{j=1} \left| a_{lj} - b_{lj} \right|<\epsilon)}  }  \right \}$,\\
\quad otherwise set $(\alpha_1^{(m+1)}, \dots, \alpha_5^{(m+1)}) = (\alpha_1^{(m)}, \dots, \alpha_5^{(m)})$\\
7. Set $m = m+1$ \\
Continue until $m=2e6$ iterations.\\
\hline 
\end{tabular}
\end{table}

\subsection{Simulation results} \label{sec:data results}

\begin{table}[htb]
\centering
\begin{tabular}{cccccc} 
 & $\hat{\alpha}_{1}$    & $\hat{\alpha}_{2}$     & $\hat{\alpha}_{3}$    & $\hat{\alpha}_{4}$  &  $\hat{\alpha}_{5}$ \\ \hline  
ABC-AR & 0.344 (0.0045) & 0.876 (0.0084) & 0.0055 (0.0012) & 0.012 (0.0025) & 4.41 (0.045) \\ 
ABC-MH & 0.351 (0.0022) & 0.891 (0.0042) & 0.0044 (0.0005) & 0.010 (0.0011) & 4.46 (0.021) \\  
&&&&&\\
& $\hat{\alpha}_{b}$    & $\hat{\beta}_{b}$     & $\hat{\alpha}_{e}$    & $\hat{\beta}_{e}$  &  $r$  \\ \hline  
ABC-AR & 0.349 (0.0044) & 4.42 (0.045) & 0.888 (0.0085) & 4.42 (0.045)  & 0.119 \\ 
ABC-MH & 0.356 (0.0022) & 4.47 (0.021) & 0.901 (0.0042) & 4.47 (0.021) & 0.120 \\
D\&H & 0.357 & 4.46 & 0.859 & 3.96  & 0.430 \\ 
\end{tabular}
\caption{Comparison of results from ABC-AR, ABC-MH, and \cite{danaher2005bacon} for bacon and eggs data.} \label{tab:results}
\end{table}

Table~\ref{tab:results} summarizes the simulation results with estimated posterior means and correlations (with standard errors).  We can see each of the three analyses yields similar results.  The ABC-AR algorithm required 399,879 proposals to obtain 500 acceptances.  The $2e6$ iterations in the ABC-MH algorithm resulted in 2,721 moves in the chain.  Given the similarity of the results, it appears the ABC-AR is performing just as well with less computational effort.

Table~\ref{tab:results} also compares our results to the model proposed by \cite{danaher2005bacon}.  Note the parameter estimates are very similar with the exception of the estimated correlation.  Our model slightly underestimates the observed table correlation of 0.23, but provides a better fit than that of \cite{danaher2005bacon}.  We can also see the estimates of $\alpha_3$ and $\alpha_4$ are near 0, suggesting the 3 parameter model of \cite{olkin2003bivariate} may be more appropriate.  

We also considered $\epsilon=80$, $\epsilon=60$, and $\epsilon=40$, but have not included the results here.  In short, the inferences were unchanged while the computational burden increased dramatically.  This was especially true in the case of $\epsilon=40$ where obtaining 500 acceptances via ABC-AR would require at least 1e8 proposals.  

\begin{table}[htb]
\centering
\begin{tabular}{c|ccccc|c}
\multicolumn {7}{c} {Eggs} \\
 Bacon       & 0    & 1     & 2    & 3     & 4    & Total \\ \hline  
0  & 250.82 & 115.11 & 48.11 & 17.17 & 4.02 & 435.23 \\ 
1  & 41.66 & 21.85 & 10.32 & 3.92 & 0.98 & 78.73 \\ 
2  & 12.52 & 6.85 &  3.66 &  1.52  & 0.46 & 25.01 \\ 
3  & 3.38 &  1.95 &  1.33 &  0.51 &  0.23 & 7.40 \\ 
4  & 0.66 &  0.45 & 0.30  & 0.17 &  0.07 & 1.65 \\ 
\hline
Total  & 309.04 & 146.21 & 63.72 & 23.29 & 5.76 & 548
\end{tabular}
\caption{Average cell counts based on the 500 accepted parameter values of the ABC-AR algorithm.} \label{tab:bias}
\end{table}

Table~\ref{tab:bias} shows the observed average cell counts for accepted tables $\mathcal{T}^\prime$ using ABC-AR.  Comparing Table~\ref{tab:data} to Table~\ref{tab:bias}, we can see no apparent pattern of bias.  We see a slight reduction in bias as $\epsilon$ is decreased from 100 to 40 (results not shown).

\subsection{Alternate analysis} \label{sec:other}

\begin{table}[htb]
\centering
\begin{tabular}{c|ccccc|c}

\multicolumn {7}{c} {Eggs$^\text{c}$} \\
 Bacon       & 0    & 1     & 2    & 3     & 4    & Total \\ \hline  
0  & 6 & 13 & 42 & 115 & 254 & 430 \\ 
1  & 1 & 6 & 16 & 29 & 34 & 86 \\ 
2  & 1 & 3 & 3 & 8 & 8 & 23 \\ 
3  & 1 & 1 & 4 & 0 & 0 & 6 \\ 
4  & 0 & 0 & 1 & 1 & 1 & 3 \\ 
\hline
Total  & 9 & 23 & 66 & 153 & 297 & 548 \\ 
\end{tabular}
\caption{Partially transposed bacon and eggs data to illustrate a negative correlation.} \label{tab:transpose}
\end{table}

This section briefly considers an alternative analysis of the bacon and eggs data with negative correlation.  In this case, the 3 parameter model of \cite{olkin2003bivariate} would be inappropriate.  To this end, consider a partial transpose of the data as in Table~\ref{tab:transpose} where the observed table correlation is -0.23.  We apply the ABC algorithms using the same steps taken before.  Under Table~\ref{tab:transpose}, the MLEs are $\tilde{\alpha}_{b}=0.3571$, $\tilde{\beta}_{b}=4.4552$, $\tilde{\alpha}_{e}=3.9593$, and $\tilde{\beta}_{e}=0.8592$.  Linking the bivariate beta model, we have
\begin{align}
{\widetilde{\alpha_1+\alpha_3}}={\tilde{\alpha}_{b}}=0.3571 \hspace{1cm} &
{\widetilde{\alpha_4+\alpha_5}}={\tilde{\beta}_{b}}=4.4552 \notag \\
{\widetilde{\alpha_2+\alpha_4}}={\tilde{\alpha}_{e}}=3.9593 \hspace{1cm} &
{\widetilde{\alpha_3+\alpha_5}}={\tilde{\beta}_{e}}=0.8592 \; . \label{eq:constraints1}
\end{align}
Prior means were chosen close to the constraints in~\eqref{eq:constraints1} subject to a correlation of -0.30.  Thus, the prior means are, $\tilde{\alpha}_1=0.9173$, $\tilde{\alpha}_2=1.7502$, $\tilde{\alpha}_3=0.8462$, $\tilde{\alpha}_4=1.1421$, and $\tilde{\alpha}_5=0.4852$, where the Monte Carlo estimate of the correlation is -0.3002. 

As with before, we ran the ABC-AR and ABC-MH algorithms for $\epsilon=100$. The results are summarized in Table~\ref{tab:results tran} where we see that they are similar for both methods.  Table~\ref{tab:bias alt} shows an observed average cell counts for accepted tables $\mathcal{T}^\prime$ using ABC-AR.  Comparing  Table~\ref{tab:transpose} to Table~\ref{tab:bias alt}, we can see there is no apparent pattern of bias. Again, we only see a slight reduction in bias as $\epsilon$ is decreased from 100 to 40, confirming that a larger $\epsilon$ suffices. 

\begin{table}[htb]
\centering
\begin{tabular}{ccccccc}
 & $\hat{\alpha}_{1}$    & $\hat{\alpha}_{2}$     & $\hat{\alpha}_{3}$    & $\hat{\alpha}_{4}$  &  $\hat{\alpha}_{5}$ & $r$ \\ \hline  
ABC-AR & 0.125 (0.0047) & 1.83 (0.043) & 0.171 (0.0053) & 2.76 (0.057) & 0.753 (0.011) & -0.246 \\
ABC-MH & 0.124 (0.0024) & 1.90 (0.023)   & 0.172 (0.0025) & 2.73 (0.027)  & 0.748 (0.005) &  -0.247 \\
\end{tabular}
\caption{Comparison of results from ABC-AR and ABC-MH algorithms for partially transposed bacon and eggs data.} \label{tab:results tran}
\end{table}

\begin{table}[htb]
\centering
\begin{tabular}{c|ccccc|c}
\multicolumn {7}{c} {Eggs$^\text{c}$} \\
 Bacon       & 0    & 1     & 2    & 3     & 4    & Total \\ \hline  
0  & 3.84 & 15.83 & 46.02 & 113.11 & 253.25 & 432.05 \\ 
1  & 0.93 &  3.86 & 11.04 & 23.29 & 37.06 & 76.18 \\ 
2  & 0.51 & 1.86 &  4.45 &  8.44 & 11.58 & 26.84 \\ 
3  & 0.36 &  0.99 &  2.15 &  2.80 &  3.72 & 10.02 \\ 
4  & 0.15 &  0.32 & 0.60 &  0.70  & 1.12 & 2.89 \\ 
\hline
Total  & 5.79 & 22.86 & 64.26 & 148.34 & 306.73 & 548 \\ 
\end{tabular}
\caption{Average cell counts based on the 500 accepted parameter values of the ABC-AR algorithm.} \label{tab:bias alt}
\end{table}

\section*{Acknowledgments}

We would like to thank Barry Arnold introducing us to this problem and providing valuable critique.  % We are also grateful to two anonymous referees and an associate editor whose valuable feedback helped improve this manuscript substantially.  
The second author's work is partially supported by NSF grant DMS-13-08270.

\begin{sidewaystable} \scriptsize
\centering
\begin{tabular}{c   c  c c c c c c c c c c} \toprule

\multicolumn {12}{c} {$\mathcal{B}\mathcal{B}(\alpha_1,\dots,\alpha_5)$ $=$ $(1,1,1,1,1)^\prime$, $ N=200, n=100$} \\ \hline

           &       & \multicolumn{2}{c} {MMLE}       & \multicolumn{2}{c} {$\mathcal{G}1$}        & \multicolumn{2}{c} {$\mathcal{U}1$}  & \multicolumn{2}{c} {$\mathcal{G}2$} & \multicolumn{2}{c} {$\mathcal{U}2$} \\ \hline

\multicolumn {12}{c} {}\\  \cline{1-12}

& Proposals  &               \multicolumn{2}{c} {--} & \multicolumn{2}{c} {13,603(1,710)} & \multicolumn{2}{c} {15,050(2,293)} & \multicolumn{2}{c} {15,191(3,862)} & \multicolumn{2}{c} {18,578(5,813)} \\ \hline
& Parameter &              Bias    & MSE     & Bias    & MSE     & Bias    & MSE  & Bias    & MSE & Bias    & MSE    \\ \hline  
&{$\hat{\alpha}_1$}    &0.008 & 0.291 & 0.313 & 0.107 & 0.479 & 0.242 & 1.532 & 2.393 & 1.895 & 3.645 \\ 
&{$\hat{\alpha}_2$}    &  -0.095 & 0.292 & 0.314 & 0.109 & 0.477 & 0.240 & 1.530 & 2.391 & 1.899 & 3.666 \\ 
${\epsilon}=0.6$ & {$\hat{\alpha}_3$}    &  0.081 & 0.156 & 0.329 & 0.127 & 0.520 & 0.292 & 1.474 & 2.259 & 1.893 & 3.715 \\&{$\hat{\alpha}_4$}    &  0.139 & 0.214 & 0.332 & 0.130 & 0.524 & 0.299 & 1.482 & 2.295 & 1.897 & 3.740 \\ 
&{$\hat{\alpha}_5$}    &  -0.014 & 0.233 & 0.338 & 0.131 & 0.532 & 0.302 & 1.574 & 2.551 & 2.005 & 4.110 \\
\hline 
%\multicolumn {12}{c} {}\\
%\multicolumn {12}{l} {${\epsilon}=0.4$} \\ \cline{1-1}
\multicolumn {12}{c} {}\\ \cline{1-12}
& Proposals        &       \multicolumn{2}{c} {--} & \multicolumn{2}{c} {50,369(10,637)} & \multicolumn{2}{c} {61,438(20,344)} & \multicolumn{2}{c} {166,248(168,771)} & \multicolumn{2}{c} {377,764(471,901)} \\ \hline
& Parameter           & Bias    & MSE     & Bias    & MSE     & Bias    & MSE  & Bias    & MSE & Bias    & MSE    \\ \hline  
&{$\hat{\alpha}_1$} & 0.008 & 0.291 & 0.290 & 0.102 & 0.464 & 0.236 & 1.132 & 1.399 & 1.365 & 2.075 \\ 
&{$\hat{\alpha}_2$} &  -0.095 & 0.292 & 0.288 & 0.100 & 0.463 & 0.236 & 1.133 & 1.400 & 1.365 & 2.080 \\ 
${\epsilon}=0.4$ & {$\hat{\alpha}_3$} &   0.081 & 0.156 & 0.270 & 0.097 & 0.456 & 0.244 & 0.995 & 1.112 & 1.207 & 1.701 \\ 
&{$\hat{\alpha}_4$} &  0.139 & 0.214 & 0.273 & 0.101 & 0.460 & 0.249 & 0.998 & 1.119 & 1.209 & 1.713 \\ 
&{$\hat{\alpha}_5$} &  -0.014 & 0.233 & 0.306 & 0.118 & 0.499 & 0.278 & 1.147 & 1.452 & 1.406 & 2.234 \\
\hline 
\multicolumn {12}{c} {}\\  \cline{1-12}
& Proposals  &               \multicolumn{2}{c} {--} & \multicolumn{2}{c} {926,581(427,404)} & \multicolumn{2}{c} {1,989,341(1,444,414)} & \multicolumn{2}{c} {10,678,542(4,732,615)} & \multicolumn{2}{c} {14,000,000(2,424,044)} \\ \hline
& Parameter &              Bias    & MSE     & Bias    & MSE     & Bias    & MSE  & Bias    & MSE & Bias    & MSE    \\ \hline  
&{$\hat{\alpha}_1$} & 0.008 & 0.291 & 0.175 & 0.065 & 0.270 & 0.130 & 0.407 & 0.237 & 0.410 & 0.306 \\ 
&{$\hat{\alpha}_2$} &  -0.095 & 0.292 & 0.152 & 0.056 & 0.243 & 0.114 & 0.384 & 0.215 & 0.364 & 0.255 \\ 
${\epsilon}=0.2$ & {$\hat{\alpha}_3$} &  0.081 & 0.156 & 0.110 & 0.037 & 0.186 & 0.075 & 0.263 & 0.113 & 0.232 & 0.110 \\ 
&{$\hat{\alpha}_4$} &  0.139 & 0.214 & 0.125 & 0.047 & 0.201 & 0.091 & 0.277 & 0.132 & 0.257 & 0.142 \\ 
&{$\hat{\alpha}_5$} &  -0.014 & 0.233 & 0.175 & 0.061 & 0.273 & 0.117 & 0.398 & 0.219 & 0.391 & 0.233 \\

\bottomrule

\end{tabular}
\caption{Bias and MSE comparisons between priors and MMLE for $A_1$.} \label{tab:5 A1}
\end{sidewaystable}

\clearpage

\begin{sidewaystable}{\centering} \scriptsize
\centering
\begin{tabular}{  c  c  c c c c c c c c c c} \toprule

\multicolumn {12}{c} {$\mathcal{B}\mathcal{B}(\alpha_1,\dots,\alpha_5)$ $=$ $(3,2.5,2,1.5,1)^\prime$, $ N=200, n=100$} \\ \hline

          &       & \multicolumn{2}{c} {MMLE}       & \multicolumn{2}{c} {$\mathcal{G}1$}        & \multicolumn{2}{c} {$\mathcal{U}1$}  & \multicolumn{2}{c} {$\mathcal{G}2$} & \multicolumn{2}{c} {$\mathcal{U}2$} \\ \hline

\multicolumn {12}{c} {}\\  \cline{1-12}

& Proposals  &              \multicolumn{2}{c} {--} & \multicolumn{2}{c} {33,995(6,940)} & \multicolumn{2}{c} {46,226(12,552)} & \multicolumn{2}{c} {26,157(5,223)} & \multicolumn{2}{c} {37,064(9,067)} \\ \hline
&Parameter &             Bias    & MSE     & Bias    & MSE     & Bias    & MSE  & Bias    & MSE & Bias    & MSE    \\ \hline  
&{$\hat{\alpha}_1$} &  0.174 & 0.833 & -1.111 & 1.243 & -1.112 & 1.242 & 0.662 & 0.466 & 0.609 & 0.384 \\ 
&{$\hat{\alpha}_2$} &    0.075 & 0.598 & -0.742 & 0.560 & -0.648 & 0.433 & 0.829 & 0.715 & 1.041 & 1.117 \\ 
${\epsilon}=0.6$ &{$\hat{\alpha}_3$} &  0.032 & 0.240 & -0.512 & 0.278 & -0.334 & 0.129 & 0.707 & 0.555 & 1.061 & 1.202 \\&{$\hat{\alpha}_4$} &    0.040 & 0.178 & -0.401 & 0.175 & -0.283 & 0.107 & 0.455 & 0.256 & 0.648 & 0.524 \\ 
&{$\hat{\alpha}_5$} &    0.047 & 0.122 & -0.227 & 0.060 & -0.244 & 0.074 & 0.374 & 0.164 & 0.328 & 0.154 \\ 
\hline
\multicolumn {12}{c} {}\\ \cline{1-12}
& Proposals  &               \multicolumn{2}{c} {--} & \multicolumn{2}{c} {137,776(34,307)} & \multicolumn{2}{c} {188,338(64,245)} & \multicolumn{2}{c} {86,778(19,653)} & \multicolumn{2}{c} {135,038(36,313)} \\ \hline
&Parameter &              Bias    & MSE     & Bias    & MSE     & Bias    & MSE  & Bias    & MSE & Bias    & MSE    \\ \hline  
&{$\hat{\alpha}_1$} & 0.174 & 0.833 & -0.981 & 0.977 & -0.999 & 1.005 & 0.714 & 0.545 & 0.646 & 0.430 \\ 
&{$\hat{\alpha}_2$} &  0.075 & 0.598 & -0.599 & 0.373 & -0.524 & 0.293 & 0.874 & 0.799 & 1.150 & 1.363 \\ 
${\epsilon}=0.4$ &{$\hat{\alpha}_3$} &  0.032 & 0.240 & -0.438 & 0.214 & -0.302 & 0.115 & 0.666 & 0.515 & 0.981 & 1.062 \\ 
&{$\hat{\alpha}_4$} &  0.040 & 0.178 & -0.370 & 0.158 & -0.288 & 0.116 & 0.395 & 0.219 & 0.521 & 0.395 \\ 
&{$\hat{\alpha}_5$} &  0.047 & 0.122 & -0.210 & 0.058 & -0.227 & 0.070 & 0.320 & 0.137 & 0.309 & 0.150 \\ 
\hline
\multicolumn {12}{c} {}\\  \cline{1-12}
& Proposals  &               \multicolumn{2}{c} {--} & \multicolumn{2}{c} {2,183,264(796,921)} & \multicolumn{2}{c} {2,702,002(1,362,655)} & \multicolumn{2}{c} {920,612(255,152)} & \multicolumn{2}{c} {1,595,734(514,779)} \\ \hline
& Parameter &              Bias    & MSE     & Bias    & MSE     & Bias    & MSE  & Bias    & MSE & Bias    & MSE    \\ \hline  
&{$\hat{\alpha}_1$} &  0.174 & 0.833 & -0.733 & 0.564 & -0.781 & 0.623 & 0.643 & 0.479 & 0.563 & 0.357 \\ 
&{$\hat{\alpha}_2$} &  0.075 & 0.598 & -0.343 & 0.149 & -0.341 & 0.148 & 0.789 & 0.689 & 1.100 & 1.275 \\ 
${\epsilon}=0.2$ &{$\hat{\alpha}_3$} &  0.032 & 0.240 & -0.261 & 0.103 & -0.200 & 0.068 & 0.542 & 0.379 & 0.801 & 0.760 \\ 
&{$\hat{\alpha}_4$} &  0.040 & 0.178 & -0.264 & 0.102 & -0.223 & 0.089 & 0.298 & 0.160 & 0.361 & 0.254 \\ 
&{$\hat{\alpha}_5$} &  0.047 & 0.122 & -0.150 & 0.044 & -0.176 & 0.055 & 0.257 & 0.112 & 0.271 & 0.130 \\ 
\bottomrule

\end{tabular}
\caption{Bias and MSE comparisons between priors and MMLE for $A_2$.} \label{tab:5 A2}
\end{sidewaystable}

\begin{sidewaystable} \scriptsize
\centering
\begin{tabular}{c   c  c c c c c c c c } \toprule

\multicolumn {10}{c} {$\mathcal{B}\mathcal{B}(\delta_1,\dots,\delta_8)$ $=$ $(2,1,1,2,4,6,2,1)^\prime$, $ N=200, n=100$} \\ \hline

           &             & \multicolumn{2}{c} {$\mathcal{G}1$}        & \multicolumn{2}{c} {$\mathcal{U}1$}  & \multicolumn{2}{c} {$\mathcal{G}2$} & \multicolumn{2}{c} {$\mathcal{U}2$} \\ \hline

\multicolumn {10}{c} {}\\ \cline{1-10}
& Proposals    & \multicolumn{2}{c} {258,167(283,773)} & \multicolumn{2}{c} {189,628(165,623)} & \multicolumn{2}{c} {205,324(223,740)} & \multicolumn{2}{c} {146,233(120,840)} \\ \hline
& Parameter     & Bias    & MSE     & Bias    & MSE  & Bias    & MSE & Bias    & MSE    \\ \hline  
& {$\hat{\alpha}_1$}    &  -0.685 & 0.471 & -0.724 & 0.529 & 0.587 & 0.358 & 0.508 & 0.273 \\ 
& {$\hat{\alpha}_2$}    &  -0.076 & 0.011 & -0.088 & 0.018 & 0.785 & 0.633 & 0.756 & 0.614 \\ 
& {$\hat{\alpha}_3$}    &  0.159 & 0.031 & 0.310 & 0.104 & 1.276 & 1.648 & 1.567 & 2.487 \\ 
${\epsilon}=0.6$ & {$\hat{\alpha}_4$}   &  -0.446 & 0.201 & -0.430 & 0.187 & 1.105 & 1.231 & 1.128 & 1.278 \\ 
& {$\hat{\alpha}_5$}    &  -2.220 & 4.970 & -2.225 & 4.967 & -0.637 & 0.552 & -0.620 & 0.449 \\ 
& {$\hat{\alpha}_6$}    &  -3.566 & 12.771 & -3.747 & 14.052 & -1.301 & 1.865 & -1.625 & 2.678 \\ 
& {$\hat{\alpha}_7$}    &  -0.990 & 0.995 & -1.006 & 1.052 & -0.029 & 0.061 & -0.057 & 0.159 \\ 
& {$\hat{\alpha}_8$}    &  -0.419 & 0.182 & -0.537 & 0.302 & 0.111 & 0.038 & -0.127 & 0.064 \\ 
\hline 
\multicolumn {10}{c} {}\\ \cline{1-10}
& Proposals          & \multicolumn{2}{c} {1,195,680(1,456,537)} & \multicolumn{2}{c} {856,819(850,336)} & \multicolumn{2}{c} {834,667(1,049,472)} & \multicolumn{2}{c} {542,178(482,364)} \\ \hline
& Parameter             & Bias    & MSE     & Bias    & MSE  & Bias    & MSE & Bias    & MSE    \\ \hline  
& {$\hat{\alpha}_1$}    &  -0.655 & 0.433 & -0.712 & 0.510 & 0.610 & 0.386 & 0.501 & 0.268 \\ 
& {$\hat{\alpha}_2$}    &  -0.098 & 0.017 & -0.111 & 0.026 & 0.709 & 0.525 & 0.669 & 0.502 \\ 
& {$\hat{\alpha}_3$}    &  0.142 & 0.029 & 0.333 & 0.123 & 1.207 & 1.488 & 1.554 & 2.460 \\ 
${\epsilon}=0.4$ & {$\hat{\alpha}_4$}   &  -0.407 & 0.168 & -0.390 & 0.154 & 1.142 & 1.315 & 1.151 & 1.332 \\ 
& {$\hat{\alpha}_5$}    &  -2.122 & 4.552 & -2.149 & 4.641 & -0.558 & 0.486 & -0.577 & 0.409 \\ 
& {$\hat{\alpha}_6$}    &  -3.412 & 11.700 & -3.647 & 13.316 & -1.152 & 1.531 & -1.548 & 2.435 \\ 
& {$\hat{\alpha}_7$}    &  -0.968 & 0.956 & -0.994 & 1.039 & -0.047 & 0.070 & -0.101 & 0.192 \\ 
& {$\hat{\alpha}_8$}    &  -0.440 & 0.202 & -0.560 & 0.328 & 0.034 & 0.029 & -0.204 & 0.087 \\ 
\hline 

\multicolumn {10}{c} {}\\  \cline{1-10}

& Proposals    & \multicolumn{2}{c} {11,893,421(3,788,548)} & \multicolumn{2}{c} {11,555,679(4,002,650)} & \multicolumn{2}{c} {6,519,920(4,768,656)} & \multicolumn{2}{c} {5,124,180(3,841,150)} \\ \hline
& Parameter      & Bias    & MSE     & Bias    & MSE  & Bias    & MSE & Bias    & MSE    \\ \hline  
& {$\hat{\alpha}_1$}    &  -0.534 & 0.291 & -0.612 & 0.378 & 0.637 & 0.421 & 0.512 & 0.280 \\ 
& {$\hat{\alpha}_2$}    &  -0.049 & 0.014 & -0.034 & 0.017 & 0.664 & 0.473 & 0.620 & 0.450 \\ 
& {$\hat{\alpha}_3$}    &  0.204 & 0.055 & 0.431 & 0.202 & 1.164 & 1.397 & 1.562 & 2.496 \\ 
${\epsilon}=0.2$ & {$\hat{\alpha}_4$}   &  -0.287 & 0.087 & -0.274 & 0.078 & 1.184 & 1.417 & 1.194 & 1.433 \\ 
& {$\hat{\alpha}_5$}    &  -1.832 & 3.419 & -1.914 & 3.694 & -0.476 & 0.419 & -0.507 & 0.339 \\ 
& {$\hat{\alpha}_6$}    &  -2.966 & 8.873 & -3.280 & 10.786 & -1.012 & 1.247 & -1.449 & 2.143 \\ 
& {$\hat{\alpha}_7$}    &  -0.799 & 0.672 & -0.822 & 0.741 & -0.037 & 0.080 & -0.109 & 0.216 \\ 
& {$\hat{\alpha}_8$}    &  -0.395 & 0.170 & -0.522 & 0.293 & -0.012 & 0.032 & -0.242 & 0.106 \\ 
\bottomrule  
\end{tabular}
\caption{Bias and MSE comparisons between priors for $A_4$.} \label{tab:8 A4}
\end{sidewaystable}

\clearpage

\setstretch{1}
\bibliographystyle{apalike}
\bibliography{ref}

\end{document}